\documentclass[twoside,11pt]{article}

\usepackage{jmlr2e}

\usepackage[dvipsnames]{xcolor}
\usepackage[colorlinks,linktoc=true,
            citecolor= Cerulean!85!green!90!black,
            linkcolor=YellowOrange,
            urlcolor=RubineRed!85!black,
            ]{hyperref}

\newcommand{\lx}[1]{{\color{black}{#1}}}
\newcommand{\yx}[1]{{\color{black}{#1}}}
\newcommand{\cm}[1]{{\color{black}{#1}}}
\newcommand{\hx}[1]{{\color{black}{#1}}}
\newcommand{\ql}[1]{{\color{black}{#1}}}
\newcommand{\wwp}[1]{{\color{black}{#1}}}
\newcommand{\yj}[1]{{\color{black}{#1}}}

\ShortHeadings{Deep Learning for Genomics: A Concise Overview}{Yue et al.}
\firstpageno{1}

\begin{document}

\title{Deep Learning for Genomics: A Concise Overview}
        
\author{\name Tianwei\ Yue$^{\dagger}$ 
    \email tianwei.vy.yue@gmail.com \\
    \addr School of Computer Science\\
        Carnegie Mellon University\\
        Pittsburgh, PA 15213, USA
       \AND
       \name Yuanxin\ Wang \email yuanxinw@alumni.cmu.edu \\
       \addr School of Computer Science\\
       Carnegie Mellon University\\
       Pittsburgh, PA 15213, USA
       \AND
       \name Longxiang\ Zhang \email longxiaz@alumni.cmu.edu \\
       \addr School of Computer Science\\
       Carnegie Mellon University\\
       Pittsburgh, PA 15213, USA 
       \AND
       \name Chunming\ Gu \email cgu15@jhmi.edu \\
       \addr Department of Biomedical Engineering \\
       School of Medicine\\
        Johns Hopkins University\\
       Baltimore, MD 21205, USA 
       \AND
       \name Haoru\ Xue \email haorux@andrew.cmu.edu \\
       \addr The Robotics Institute\\
       Carnegie Mellon University\\
       Pittsburgh, PA 15213, USA 
       \AND
       \name Wenping\ Wang \email wenpingw@alumni.cmu.edu\\
       \addr School of Computer Science\\
       Carnegie Mellon University\\
       Pittsburgh, PA 15213, USA 
       \AND
       \name Qi\ Lyu \email lyuqi1@msu.edu\\
       \addr Department of Computational Mathematics, Science, and Engineering\\
       Michigan State University\\
       East Lansing, MI 48824, USA 
       \hfill 
    \AND
       \name Yujie\ Dun \email dunyj@mail.xjtu.edu.cn \\
       \addr School of Information and Communications Engineering\\
        Xi'an Jiaotong University\\
         Xi'an, Shaanxi 710049, China \\
      \hfill 
    \AND
      \name Corresponding\ Author: 
      \name Tianwei\ Yue 
}

\editor{}
\maketitle 

\newpage
\tableofcontents

\clearpage

\begin{abstract}
This data explosion driven by advancements in genomic research, such as high-throughput sequencing techniques, is constantly challenging conventional methods used in genomics. In parallel with the urgent demand for robust algorithms, deep learning has succeeded in various fields such as vision, speech, and text processing. Yet genomics entails unique challenges to deep learning since we expect a superhuman intelligence that explores beyond our knowledge to interpret the genome from deep learning. A powerful deep learning model should rely on insightful utilization of task-specific knowledge. In this paper, we briefly discuss the strengths of different deep learning models from a genomic perspective so as to fit each particular task with proper deep architecture, and remark on practical considerations of developing deep learning architectures for genomics. We also provide a concise review of deep learning applications in various aspects of genomic research and point out current challenges and potential research directions for future genomics applications.
\end{abstract}


\section{Introduction}

Even since \cite{1stDNAunderstandn1953} first interpreted DNA molecules as the physical medium carrying genetic information, human beings have been striving to gather biological data and decipher the biological processes guided by genetic information. 
By the time of 2001, the Human Genome Project launched in 1990 had drafted the raw information of a typical human genome \citep{initialHG2001}. Many other genome projects, including 
FANTOM \citep{FANTOM2001functional},
ENCODE \citep{encode2012integrated},
Roadmap Epigenomics \citep{Roadmap2015integrative}, were also launched in succession. These collaborative efforts made an abundance of DNA data available and thus allowed a global perspective on the genome of different species, leading to the prosperity of genomic research.

Genomic research aims to understand the genomes of different species. It studies the roles assumed by multiple genetic factors and the way they interact with the surrounding environment under different conditions. In contrast to genetics which deals with a limited number of specific genes, genomics takes a global view that involves the entirety of genes possessed by an organism \citep{jax}. For example, a study of homo sapiens involves searching through
approximately 3 billion units of DNA, 
containing protein-coding genes, RNA genes, \textit{cis}-regulatory elements, long-range regulatory element, and transposable elements \citep{bae2015genetic}.
Additionally, genomics is becoming increasingly data-intensive with the advancement in genomic research, such as the cost-effective next-generation sequencing technology that produces the entire readout of the DNA of an organism. This high-throughput technology is made available by more than 1,000 sequencing centers cataloged by OmicsMaps (\url{http://omicsmaps.com/}) on nearly every continent \citep{stephens2015big}. The vast trove of information generated by genomic research provides a potential exhaustive resource for scientific study with statistical methods. These statistical methods can be used to identify different types of genomic elements, 
such as exons, introns, promoters, enhancers, positioned nucleosomes, splice sites, untranslated region (UTR), \textit{etc}. In addition to recognizing these patterns in DNA sequences, models can take other genetic and genomic information as input to build systems to help understand the biological mechanisms of underlying genes. A large variety of data types are available, such as chromatin accessibility assays (\textit{e.g.} MNase-seq, DNase-seq, FAIRE), genomic assays (\textit{e.g.} microarray, RNA-seq expression), transcription factor (TF) binding ChIP-seq data, gene expression profiles, and histone modifications, \textit{etc} \citep{libbrecht2016understanding}. Most of these data are available through portals like GDC (\url{https://portal.gdc.cancer.gov/}), dbGaP (\url{https://www.ncbi.nlm.nih.gov/gap}), GEO (\url{https://www.ncbi.nlm.nih.gov/geo/}), just to name a few. A combination of various data can bring about deeper insights into genes so as to help researchers to locate the information of interest. 

On the other hand, the development of deep learning methods has granted the computational power to resolve these complex research questions \citep{lecun2015deep,wang2017origin}. Its success has already been demonstrated by the revolutionizing achievements in the field of artificial intelligence, \textit{e.g.}, image recognition, object detection, audio recognition, and natural language processing, \textit{etc.}
The boom of deep learning is supported by the successive introduction of a variety of deep architectures, including autoencoders \citep{ae1975cognitron} and their variants,
multilayer perceptron \citep[MLP;][]{mlp1985bp,mlp1997introduction}, restricted Boltzmann machines \citep[RBMs;][]{rbm1986hinton},
deep belief networks \citep[DBNs;][]{dbn2006intro}, 
convolutional neural networks \citep[CNN;][]{fukushima1982neocognitronCNN,cnn1990lecun}, recurrent neural networks \citep[RNN;][]{elman1990findingRNN}, 
Long Short-Term Memory \citep[LSTM;][]{lstm1997}, 
\yj{
Transformers \citep{attention-is-all-you-need}, 
Large Language Models \citep{gpt-1, gpt-2, gpt-3}
},
and other recently appearing architectures that will be introduced later in this article. The strong flexibility and high accuracy of deep learning methods guarantee them sweeping superiority over other existing methods on these classical tasks.


The intersection of deep learning methods and genomic research may lead to a profound understanding of genomics that will benefit multiple fields including precision medicine \citep{leung2016machine}, pharmacy (\textit{i.e.} drug design), and even agriculture, \textit{etc}. Take medicine for example, medical research and its applications such as gene therapies, molecular diagnostics, and personalized medicine could be revolutionized by tailoring high-performance computing methods to analyzing available genomic datasets. Also, the process of developing new drugs takes a long period and is usually very costly. To save time and cost, the general approach taken by pharmaceutical companies is to try to match the candidate protein identified by researchers with their known drug molecules. \citep{pharmaceutical}. \yx{As we are facing larger-scale and more complex medical demands, cutting-edge deep learning techniques such as large language models show emergent capabilities to efficiently and effectively deal with unprecedented challenges such as Covid-19 \cite{hammad_hybrid_2023, genslm}.} All these benefits indicate the necessity of utilizing powerful and specially designed deep learning methods to foster the development of the genomics industry. This article aims to offer a concise overview of the current deep learning applications in genomic research, and, if possible, point out promising directions for further applying deep learning in the genomic study.

The rest of this article is organized as follows: we first briefly introduce the genomic study powered by deep learning characterized by deep learning architectures in Section~\ref{sec:architecture}, with additional discussions offered in Section~\ref{sec:insights}. Then we discuss the use of deep learning methods on various topics in genomics in Section~\ref{sec:applications}, followed by our summarization of the current challenges and potential research directions in Section~\ref{sec:opportunities}. Finally, conclusions are drawn in Section~\ref{sec:conclusion}.




\section{Deep Learning Architectures: A Genomic Perspective}
\label{sec:architecture}

Various deep learning algorithms have their own advantages to resolve particular types of problems in genomic applications. For example, CNNs that are famous for capturing features in image classification tasks have been widely adopted to automatically learn local and global characterization of genomic data. RNNs that succeed in speech recognition problems are skillful at handling sequence data and thus were mostly used to deal with DNA sequences. Autoencoders are popular for both pre-training models and denoising or pre-processing the input data.
When designing deep learning models, researchers could take advantage of these merits to efficiently extract reliable features and reasonably model the biological process. This section will review some details on each type of deep architecture, focusing on how each of their advantages can benefit the specific genomic research questions. This article will not cover the standard introduction of deep learning methods, readers can visit classical textbooks \cite[\textit{e.g.}][]{goodfellow2016deep} or concise tutorials \citep[\textit{e.g.}][]{wang2017origin} if necessary.


\subsection{Convolutional Neural Networks}
\label{cnn-section}
Convolutional neural networks (CNNs) are one of the most successful deep learning models for image processing owing to their outstanding capacity to analyze spatial information. Early applications of CNNs in genomics relied on the fundamental building blocks of CNNs in computer vision \citep{cnn2012imagenet} to extract features. \citet{cnnIntro2016} described the adaptation of CNNs from the field of computer vision to genomics as accomplished by comprehending 
a window of genome sequence as an image. 

The highlight of CNNs is the dexterity of automatically performing adaptive feature extraction during the training process.
For instance, CNNs can be applied to discover meaningful recurring patterns with small variances, such as genomic sequence motifs. This makes CNNs suitable for motif identification and therefore binding classification \citep{deepmotif2016}.

Recently CNNs have been shown to take a lead among current algorithms for solving several sequence-based problems. \citet[DeepBind]{DeepBind} and \citep{cnnIntro2016} successfully applied CNNs to model the sequence specificity of protein binding. \citet[DeepSEA]{deepsea2015predicting} developed a conventional three-layer CNN model to predict from only genomic sequence the effects of non-coding variants. \citet[Basset]{Basset2016Kelley} adopted a similar architecture to study the functional activities of DNA sequence.

Though multiple research has demonstrated the superiority of CNNs over other existing methods, inappropriate structure design would still result in even poorer performance than conventional models \cite{cnnIntro2016}. Therefore, what lies in the center for researchers to master and optimize the ability of CNNs is to skillfully match a CNN architecture to each particular given task. To achieve this, researchers should have an in-depth understanding of CNN architectures as well as take into consideration of biological background. \citet{cnnIntro2016} developed a parameterized convolutional neural network to conduct a systematic exploration of CNNs on two classification tasks, motif discovery, and motif occupancy. They performed a hyper-parameter search using Mri (\url{https://github.com/Mri-monitoring/Mri-docs/blob/master/mriapp.rst}) and mainly
examined the performance of nine variants of CNNs, and concluded that CNNs are not necessary to be deep for motif discovery task as long as the structure is appropriately designed. When applying CNNs in genomic, since deep learning models are always over-parameterized, simply changing the network depth would not account for much improvement in model performance. \yx{On this direction, \citet{xuan2019dual} designed a dual CNN with attention mechanisms to extract deeper and more complex feature representations of lncRNA (long non-coding RNA genes); while Kelley \textit{et al.} \cite{basenji1, basenji2} took a different path in using dilated convolution instead of classical convolution to share information across long distances without adding depth indefinitely.}

\subsection{Recurrent Neural Networks}
\label{rnn-section}
Recurrent neural networks (RNNs) raised a surge of interest owning to its impressive performance on sequential prediction problems such as language translation, summarization, and speech recognition. RNNs outperform CNNs and other early deep neural networks (DNNs) on sequential data thanks to their capability of processing long ordered sequences and memorizing long-range information through recurrent loops. Specifically, RNNs scan the input sequences sequentially and feed both the previously hidden layer and current input segment as the model input so that the final output implicitly integrates both current and previous information in the sequence. \citet{brnn1997} later proposed bidirectional RNN (BRNN) for use cases where both past and future contexts in the input matter. 

The cyclic structure makes a seemingly shallow RNN over long-time prediction actually very deep if unrolled in time. To resolve the problem of vanishing gradient rendered by this, \citet{lstm1997} substituted the hidden units in RNNs with LSTM units to truncate the gradient propagation. \citet{gru2014learning} introduced Gated Recurrent Units (GRUs) with a similar proposal.

Genomics data are typically sequential and often considered languages of biological nature. Recurrent models are thus applicable in many scenarios. For example, \citet[ProLanGO]{cao2017prolango} built an LSTM-based Neural Machine Translation, which converts the task of protein function prediction to a language translation problem by interpreting protein sequences as the language of Gene Ontology terms. \citet{DeepNano2017} developed DeepNano for base calling, \citet{DanQ2016} proposed DanQ to quantify the function of non-coding DNA, \citet{sonderby2015convolutionalLSTM} devised a convolutional LSTM to predict protein subcellular localization from protein sequences, and \citet{busia2016protein} applied the idea of seq-to-seq learning to their model for protein secondary structure prediction conditioned on previously predicted labels. \yx{Furthermore, sequence-to-sequence learning for genomics is boosted by attention mechanisms: \citet{lstm-attention} introduced an attention-based approach where a hierarchy of multiple LSTM modules are used to encode input signals and model how various chromatin marks cooperate; similarly, \citet{rnn-attention} used LSTM as feature extractor and attention modules as importance scoring functions to identify regions of the RNA sequence that bind to proteins.}

\subsection{Autoencoders}

Autoencoders, conventionally used as pre-processing tools to initialize the network weights, have been extended to stacked autoencoders \cite[SAEs;][]{bengio2007sae}, denoising autoencoders \cite[DAs;][]{sae2008intro}, contractive autoencoders \cite[CAEs;][]{rifai2011cae}, \textit{etc}. Now they have proved successful in feature extraction because of being able to learn a compact representation of input through the encode-decode procedure. For example, \citet{GeneExphhw2015} applied stacked denoising autoencoders (SDAs) for gene clustering tasks. They extracted features from data by forcing the learned representation resistant to a partial corruption of the raw input. More examples can be found in Section \ref{GeneExpCha}. Besides, autoencoders are also used for dimension reduction in gene expression, \textit{e.g.} \cite{GeneExpADAGE2014,ADAGE22016,eADAGE2017}. When applying autoencoders, one should be aware that better reconstruction accuracy does not necessarily lead to model improvement \citep{rampasek2017dr}.

Variational Autoencoders (VAEs), though named "autoencoders", was rather developed as an approximate-inference method to model latent variables. Based on the structure of autoencoders, \citet{kingma2013auto} added stochasticity to the encoded units and added a penalty term encouraging the latent variables to produce a valid decoding. VAEs aim to deal with the problems of which each data has a corresponding latent representation and are thus useful for genomic data, among which there are complex interdependencies. \citet{rampasek2017dr} presented a two-step VAE-based model for drug response prediction, which first predicts the post- from the pre-treatment state in an unsupervised manner, then extends it to the final semi-supervised prediction. This model was based on data from Genomics of Drug Sensitivity in Cancer \cite[GDSC;][]{GDSCdata} and Cancer Cell Line Encyclopedia \citep[CCLE;][]{barretina2012cancer}. \ql{VAEs can also be used in many other genomic applications including cancer gene expression prediction \cite{way2017vaeCancer, way2017evaluating}, single cell feature extraction for unmasking tumor heterogeneity \cite{rashid2021dhaka}, metagenomic binning \cite{metagenome}, DNA methylome dataset construction \cite{methylome}, \textit{etc}.}


\subsection{Emergent Deep Architectures}
As deep learning constantly showing success in genomics, researchers are expecting deep learning higher accuracy than simply outperforming statistical or machine learning methods. To this end, the vast majority of work nowadays approached genomic problems from more advanced models beyond classic deep architectures, or employing hybrid models. Here we review some examples of recent appearing deep architectures 
by which skillfully modifying or combining classical deep learning models.

\subsubsection{Beyond Classic Models}

Most of these emergent architectures are of natural designs modified from classic deep learning models. Researchers began to leverage more genomic intuitions to fit each particular problem with a more advanced and suitable model.

Motivated by the fact that protein folding is a progressive refinement \citep{min2017deep} rather than an instantaneous process, \citet{DSTnn2012deep} designed DST-NNs for residue-residue contact prediction. It consists of a 3D stack of neural networks in which topological structures (same input, hidden, and output layer sizes) are identical in each stack. Each level of this stacked network can be regarded as a distinct contact predictor and can be trained in a supervised matter to refine the predictions of the previous level, hence addressing the typical problem of vanishing gradients in deep architectures. The spatial features in this deep spatiotemporal architecture refer to the original model inputs, while temporal features are gradually altered so as to progress to the upper layers. \citet[DeepCpG]{DeepCpG2017Angermueller}
took advantage of two CNN sub-models and a fusion module to predict DNA methylation states. The two CNN sub-models take different inputs and thus focus on disparate purposes. The CpG module accounts for correlations between CpG sites within and across cells, while the DNA module detects informative sequence patterns (motifs). Then the fusion module can integrate higher-level features derived from two low-level modules to make predictions. Instead of subtle modifications or combinations, some works focused on depth trying to improve the model performance by designing even deeper architectures. \citet{ultraDeep2017} developed an ultra-DNN consisting of two deep residual neural networks 
to predict protein contacts from a sequence of amino acids. Each of the two residual nets in this model has its particular function. A series of 1D convolutional transformations are designed for extracting sequential features (\textit{e.g.}, sequence profile, predicted secondary structure, and solvent accessibility). The 1D output is converted to a 2D matrix by an operation similar to the outer product and merged with pairwise features (\textit{e.g.}, pairwise contact, co-evolution information, and distance potential). Then they are together fed into the second residual network, which consists of a series of 2D convolutional transformations. The combination of these two disparate residual nets makes possible a novel approach to integrate sequential features and pairwise features in one model.

\subsubsection{Hybrid Architectures}

The fact that each type of DNN has its own strength inspires researchers to develop hybrid architectures that could well utilize the potential of multiple deep learning architectures. DanQ \citep{DanQ2016} is a hybrid convolutional and recurrent DNN for predicting the function of non-coding DNA directly from sequence alone. A DNA sequence is input as the one-hot representation of four bases to a simple convolutional neural network with the purpose of scanning motif sites. Motivated by the fact that the motifs can be determined to some extent by the spatial arrangements and frequencies of combinations of DNA sequences \citep{DanQ2016}, the purported motifs learned by CNN are then fed into a BLSTM. Similar convolutional-recurrent designs were further discussed by \citet[Deep GDashboard]{Dashboard}. They demonstrated how to understand three deep architectures: convolutional, recurrent, and convolutional-recurrent networks, and verified the validity of the features generated automatically by the model through visualization techniques. They argued that a CNN-RNN architecture outperforms CNN or RNN alone based on their experimental results on a transcription factor binding site (TFBS) classification task. The feature visualization achieved by Deep GDashboard indicated that CNN-RNN architecture is able to model both motifs as well as dependencies among them. \citet{sonderby2015convolutionalLSTM} added a convolutional layer between the raw data and LSTM input to address the problem of protein sorting or subcellular localization. In total, there are three types of models proposed and compared in the paper, a vanilla LSTM, an LSTM with an attention model used in a hidden layer, and an ensemble of ten vanilla LSTMs. They achieved higher accuracy than previous benchmark models in predicting the subcellular location of proteins from DNA sequences while no human-engineered features were involved. \citet{almagro2017deeploc} proposed a hybrid integration of RNN, BLSTM, attention mechanism, and a fully connected layer for protein subcellular localization prediction; each of the four modules is designed for a specific purpose. These hybrid models are increasingly favored by recent research, \textit{e.g.} \cite{singh2016EnhancerPromoter}.


\yx{
\subsection{Transformer-based Large Language Models}
\label{llm-architecture}
As mentioned in Section \ref{cnn-section} and \ref{rnn-section}, many prior deep-learning works utilized CNNs and RNNs to solve genomics tasks. However, there are several intrinsic limitations of these two architectures: (1) CNNs might fail to capture the global understanding of a long DNA sequence due to its limited receptive field (2) RNNs could have difficulty in capturing useful long-term dependencies because of vanishing gradients and suffer from low-efficiency problem due to its non-parallel sequence processing nature (3) Both architectures need extensive labeled data to train. These limitations hinder them from coping with harder genomics problems since these tasks usually require the model to (1) understand long-range interactions (2) process very long sequences efficiently (3) perform well even for low-resource training labels. 

Transformer-based \cite{attention-is-all-you-need} language models such as BERT \cite{bert} and GPT family \cite{gpt-1, gpt-2, gpt-3}, then become a natural fit to overcome these limitations. Their built-in attention mechanism learns better representations that can be generalized to data-scarce tasks via larger receptive fields. \cite{benegas2022dna} found that a pre-trained large DNA language model is able to make accurate zero-shot predictions of non-coding variant effects. Similarly, according to \cite{nucleotide-transformer}, these language model architectures generate robust contextualized embeddings on top of nucleotide sequences and achieve accurate molecular phenotype prediction even in low-data settings.

Instead of processing input tokens one by one sequentially as RNNs, transformers process all input tokens more efficiently at the same time in parallel. However, simply increasing the input context window infinitely is infeasible since the computation time and memory scale quadratically with context length in the attention layers. Several improvements have been made from different perspectives: \citet{hyenaDNA} uses the Hyena architecture \cite{hyenaModel} and scales sub-quadratically in context length, while \citet{dnabert-2} replace k-mer tokenization used in \citet{dnabert} with Byte Pair Encoding (BPE) to achieve a 3x efficiency improvement.

In light of dealing with extremely long-range interactions in DNA sequences, the Enformer model \cite{enformer} employs transformer modules that scale a five times larger receptive field compared to previous CNN-based approaches \cite{expecto, basenji1, basenji2}, and is capable of detecting sequence elements that are 100 kb away.  Moreover, the recent success of ChatGPT \cite{chatgpt} and GPT-4 \cite{gpt-4} further illustrated the emergent capabilities of large language models (LLMs) to deal with such long DNA sequences. A typical transformer-based genomics foundational model can only take 512 to 4k tokens as input context, which is less than 0.001\% of the human genome. \citet{hyenaDNA} proposed an LLM-based genomic model that expands the input context length to 1 million tokens at the single nucleotide level, which is up to 500x increase over previous dense attention-based models. 

}

\section{Deep Learning Architectures: Insights and Remarks}
\label{sec:insights}

Applications of deep learning in genomic problems have fully proven their power. Although the pragmatism of deep learning is surprisingly successful, this method suffers from lacking the physical transparency to be well interpreted so as to better assist the understanding of genomic problems. What is auspicious in genomic research is that researchers have done lots of work to visualize and interpret their deep learning models. Besides, it is also constructive to take into additional considerations beyond the choice of deep learning architectures. In this section, we review some visualization techniques that bring about insights into deep learning architectures and add remarks on model design that might be conducive to real-world applications.

\subsection{Model Interpretation}

People expect deep networks to succeed not only in predicting results but also in identifying meaningful DNA sequence signals and giving further insights into the problems being solved. The interpretability of a model appears to be crucial when it comes to application. However, the technology of deep learning has exploded not only in prediction accuracy but also in complexity as well. Connections among network units so are convoluted that the
information is widespread throughout the network and thus perplexing to be captured \citep{openBox}.
People are carrying out efforts to remedy this pitfall since prediction accuracy alone does not guarantee that deep architectures are a better choice over traditional statistical or machine learning methods in applications. \yx{Different visualizations techniques are also being actively developed \cite{deepenhancer2016,DeepChrome2016,mikolov2013distributed,sonderby2015convolutionalLSTM,riesselman2017deepGenerative}. }

The image classification field is where people started deciphering deep networks. \citet{zeiler2014visualizing} gave insights into the function of intermediate features by mapping hidden layers back to the input through deconvolution, a technique described in that paper. \citet{simonyan2013deep} linearly approximate the network by first-order Taylor expansion and obtained Saliency Maps from a ConvNet by projecting back from the dense layers of the network. People also searched for an understanding of genes through deep networks. \citet{denas2013deep} managed to pass the model knowledge back into the input space through the inverse of the activation function, so that biologically-meaningful patterns can be highlighted. \citet[Dashboard]{Dashboard} adopted Saliency Maps to measure nucleotide importance. Their work provided a series of visualization techniques to detect motifs, or sequence patterns from deep learning models, 
and went further to discuss the features extracted by CNNs and RNNs. Similarly, \citet{DeepBind} visualized the sequence specificities determined by DeepBind through mutation maps that indicate the effect of variations on bound sequences. Note that works conducted appropriately by classic models do not need additional techniques to visualize features, \textit{e.g.} \citet{parnamaa2017Subcellular} trained an 11-layer CNN for prediction protein subcellular localization from microscopy images, and easily interpreted their model by features at different layers. 

\yx{The rise of attention-based models also opened up new avenues for interpretability in genomics. \citet{lstm-attention} argued that attention scores provide a better interpretation than traditional feature visualization methods such as Saliency maps. According to \citet{chen2019interpretable}, the visualization of attention weight changes can be used to understand when binding signal peaks move along the genomic sequence. \citet{ghotra2021designing} further emphasized the importance of the convolutional layer(s) learning identifiable motifs for the attention maps to be interpretable.}


\subsection{Transfer Learning and Multitask Learning}
\label{sub-sec:transfer}

The concept of transfer learning is naturally motivated by human intelligence that people can apply the knowledge that has already been acquired to address newly encountered problems. Transfer learning is such a framework that allows deep learning to adapt the previously-trained model to exploit a new but relevant problem more effectively \citep{lu2015transfer}. It has been successfully applied to other fields, such as language processing \citep{cirecsan2012transfer} or audio-visual recognition \citep{moon2014multimodal}. Readers could find 
surveys on transfer learning by \citet{transfer2010survey} or \citet{weiss2016survey}. Additionally, multitask learning is an approach that inductively shares knowledge among multiple tasks. By learning related tasks in parallel while using shared architectures, what is learned by a single task can be auxiliary to those related. An overview of multitask learning, which especially focuses on neural networks, can be found in 
\citet{ruder2017overview}. \citet{widmer2012multitask} briefly discussed multitask learning from a biological perspective.

Early adaptation of transfer learning in genomics was based on machine learning methods such as SVMs \citep{schweikert2009empirical,mei2013probability,xu2011survey}. Recent works have also involved deep learning. For example, \citet{transferBioIma2016deep} developed a CNN model to analyze gene expression images for automatic controlled vocabulary (CV) term annotation. They pre-trained their model on ImageNet (\url{http://www.image-net.org/}) to extract general features at different scales, then fine-tuned the model by multitask learning to capture CV term-specific discriminative information. \citet{liu2016pedla} developed an iterative PEDLA to predict enhancers across multiple human cells and tissues. 
They first pre-trained PEDLA on data derived from any cell type/tissue an unsupervised manner, then iteratively fine-tuned the model on a subsequent cell type/tissue supervisedly, using the trained model of the previous cell type/tissue as initialization.
\citet{cohn2018enhancer} transferred deep CNN parameters between networks trained on different species/datasets for enhancer identification. 
\citet[][TFImpute]{qin2017imputation} adopted a CNN-based multi-task learning setting to borrow information across TFs and cell lines to predict cell-specific TF binding for TF-cell line combinations from only a small portion of available ChIP-seq data. They were able to predict TFs in new cell types by models trained unsupervisedly on TFs where ChIP-seq data are available, which took the right step in the direction of developing a domain transfer model across cell types. \citet{qi2010semi} proposed a semi-supervised multi-task framework for protein-protein interaction (PPI) predictions. 
They applied the MLP classifier trained supervisedly to perform an auxiliary task that leverages partially labeled examples. The loss of the auxiliary task is added to MLP loss so that the two tasks can be jointly optimized. \citet{wang2017extracting} worked on the same problem by introducing a multi-task convolutional network model for representation learning. \citet{deepsea2015predicting} incorporated multitask approach for noncoding-variant effects prediction on chromatin by jointly learning across diverse chromatin factors. \ql{\citet{multi-task-multi-modal} proposed a task relationship learning framework to automatically investigate the inherent correlation between diagnosis and prognosis genomics tasks, while DeepND \cite{beyreli2022deepnd} claimed to learn the shared and disorder-specific features using multitask learning setting where several tasks are solved together.}


\subsection{Multi-view Learning}
\label{sub-sec:multi-modal}
As the current technology has made available data from multi-platform or multi-view inputs with heterogeneous feature sets, multi-view deep learning appears to be an encouraging direction for future deep learning research which exploits information across the datasets, capturing their high-level associations for prediction, clustering as well as handling incomplete data. Readers can visit \citet{li2016multiviewL} for a survey on multi-view methods if interested. In many applications, we can approach the same problem from different types of data, such as in computer visions when audio and video data are both available \citep{sound2005pixels,wang2017select}. Genomics is an area where data of various types can be assimilated naturally. For example, abundant types of genomic data (\textit{e.g.}, DNA methylation, gene expression, miRNA expression data) for the same set of tumor samples have been made available by the state-of-the-art high-throughput sequencing technologies \citep{GeneExpMultimodal2015}. Therefore, it is natural to think of leveraging multi-view information in genomics to achieve a better prediction than that of a single view. \citet{gligorijevic2015methods} and \citet{li2016review} reviewed some methods for multi-view biological data integration with instructive considerations. 

Multi-view learning can be achieved by, for example, concatenating features, ensemble methods, or multi-modal learning (selecting specific deep networks, as sub-networks of the main model, for each view, then integrating them in higher layers), 
just to name a few. The previously mentioned ultra-DNN \citep{ultraDeep2017} is a case in point, where it adopted 1D and 2D CNNs respectively for sequential features and spatial features. \citet{GeneExpMultimodal2015} proposed a multi-modal DBN to integrate gene expression, DNA methylation, miRNA, and drug response data to cluster cancer patients and define cancer subtypes.
Their stacked Gaussian-restricted Boltzmann machines (gRBM) are trained by contrastive divergence, different modalities are integrated via stacking hidden layers, and common features are effused from inherent features derived from multiple single modalities. \yx{On this direction, \citet{zhang2015deep} utilized a multimodal DBN framework to integrate RNA primary sequence, predicted secondary, and tertiary structures, while a later work \citet{pan2017rna} went a step further with a hybrid multimodal framework combining CNNs and DBNs to predict RNA-binding protein interaction sites and motifs on RNAs using 5 different modalities. Additionally, instead of dealing with different modalities separately, some research started to explore multi-modal interactions: for example, \citet{multi-task-multi-modal} performed integrative analysis on histopathological image and genomic data for cancer diagnosis and prognosis, while \citet{10.1093/bioinformatics/btab185} explored both intra-modality and inter-modality feature modules for genomic data and pathological images.}

\section{Genomic Applications}
\label{sec:applications}

In this section, we review several aspects of genomic problems that can 
be approached from deep learning methods and discuss how deep learning moves forward in these fields.

\subsection{Gene expression}
Gene expression is a highly regulated process by which the genetic instructions in DNA are converted into functional products such as proteins and other molecules, and also respond to the changing environment accordingly. Namely, genes encode protein synthesis, and self-regulate the functions of the cell by adjusting the amount and type of proteins it produces \citep{GenExp}.
Here we review some research that applied deep learning to analyze how gene expression is regulated.

\subsubsection{Gene expression Characterization}\label{GeneExpCha}

%
%
%
An increasing number of genome-wide gene expression assays for different species have become available in public databases, \textit{e.g.} the Connectivity Map (CMap) project was launched to create a reference collection of gene expression profiles that can be used to identify functionally connected molecules \citep{CMap2006}, these databases greatly facilitated the computational models for biological interpretation of these data. At the same time, recent works have suggested better performance obtained by deep learning models on gene expression data; \citet{urda2017deep} used a deep learning approach to outperform LASSO in analyzing RNA-Seq gene expression profile data. 

The empirical results of early works that applied principal component analysis (PCA) on gene expression data to capture cluster structure have shown that this mathematical tool was not effective enough to allow some complicated biological considerations \citep{GeneExpPCA2001}. 
Also, since the reliability of the cross-experiment datasets is limited by technical noise and unmatched experiment conditions \citep{eADAGE2017}, researchers are considering the denoising and enhancement of the available data instead of directly finding principal components.

Denoising autoencoders came in handy since they do not merely retain the information of raw data, but also generalize meaningful and important properties of the input distribution across all input samples. Even shallow denoising autoencoders 
can be proven effective in extracting biological insights. \citet{danaee2017deep} adopted SDAs to detect functional features in breast cancer from gene expression profile data. \citet[ADAGE]{GeneExpADAGE2014} presented an unsupervised approach that effectively applied SDA to capture key biological principles in breast cancer data. ADAGE is an open-source project for extracting relevant patterns from large-scale gene expression datasets. 
\citet{ADAGE22016} further improved ADAGE to successfully extract both clinical and molecular features. 
%
To build better signatures that are more consistent with biological pathways and enhance model robustness, \citet{eADAGE2017} developed an ensemble ADAGE (eADAGE) to integrate stable signatures across models. These three similar works were all experimented on Pseudomonas aeruginosa gene expression data. Additionally, \citet{GeneExphhw2015} demonstrated the efficacy of using the enhanced data by multi-layer denoising autoencoders to cluster yeast expression microarrays into known modules representing cell cycle processes.
Motivated by the hierarchical organization of yeast transcriptomic machinery, \citet{GeneExpHierachical2016learning} adopted a four-layered autoencoder network with each layer accounting for a specific biological process in gene expression. This work also introduced sparsity into autoencoders. Edges of denoising autoencoders over PCA and independent component analysis (ICA) were clearly illustrated in the aforementioned works. 

Some other works moved to variational inference in autoencoders, which is assumed to be more skillful to capture the internal dependencies among data. \citet{way2017evaluating} trained VAE-based models to reveal the underlying patterns in the pathways of gene expression, and compared their three VAE architectures to other dimensionality reduction techniques, including the aforementioned ADAGE \citep{GeneExpADAGE2014}. 
\citet{dincer2018deepprofile} introduced the DeepProfile, a framework featuring VAE, to extract latent variables that are predictive for acute myeloid leukemia from expression data. \citet{sharifi2018deep} proposed Deep Genomic Signature (DGS), a pair of VAEs that are trained over unlabelled and labeled data separately from expression data for predicting metastasis. 

Another thread for utilizing deep learning to characterize gene expression is to describe the pairwise relationship. \citet{wang2017extracting} showed that CNN can be seen as an effective replacement for the frequently used Pearson correlation applied to pair of genes, therefore they built a multi-task CNN that can consider the information of GO semantics and interaction between genes together to extract higher level representations of gene pairs for the further classification task, which is further extended by two shared-parameter networks \citep{cao2017learning}. \yx{Recently LLMs have come into play for such pairwise relationship: \citet{cui2023scgpt} introduced a GPT-based foundational model and found a positive pairwise correlation between the similarity of the gene embeddings and the number of common pathways shared by these genes; similarly, \citet{yang2022scbert} utilized the attention weights in transformers to reflect the contribution of each gene and the interaction of gene pairs.}

\subsubsection{Gene expression Prediction}
 
Deep learning approaches for gene expression prediction have outperformed other existing algorithms. 
For example, \citet{geneExpressionInfer2016} presented a three-layer feed-forward neural network for gene expression prediction of selected landmark genes that achieved better performance than linear regression. This model, D-GEX, is of the multi-task setting and was tested on two types of expression data, the microarrays and RNA-Seqs. \citet{xie2017deep} showed that their deep model based on MLP and SDAs outperformed Lasso and Random Forests in predicting gene expression quantifications from SNP genotypes.

When making predictions from gene sequences, deep learning models have been shown fruitful in identifying the context-specific roles of local DNA-sequence elements, 
then the further inferred regulatory rules can be used to predict expression patterns \citep{GeneExpPredSeq2004}. Successful prediction usually relies much on the proper utilization of biological knowledge. Therefore, it could be more efficient to pre-analyze the contextual information in DNA sequences than directly making predictions. Deep learning models could refer to two early machine learning works that apply Bayesian networks to predict gene expression based on their learned motifs \citep{GeneExpPredSeq2004, GeneExpPredSeq2007re}.
%

In most applications, the powerful of deep learning algorithms is paled by biological restrictions. Therefore, instead of only using sequence information, combing epigenetic data into the model might add to the explanatory power of the model. For example, the correlation between histone modifications and gene regulation was suggested experimentally in \citet{1histone2009}, \citet{2histone2011} and \citet{3histone2013}, and has already been studied in some machine learning works before \citep{LR2010histone, SVM2011histone,RF2012histone,rule2015histone}. 
\citet{DeepChrome2016} presented DeepChrome, a unified discriminative framework stacking an MLP on top of a CNN, and achieved an average AUC of 0.8 in binary classification task that predicts high or low gene expression level. The input was separated into bins so as to discover the combinatorial interactions among different histone modification signals. 
The learned region representation is then fed into an MLP classifier that maps to gene expression levels. Additionally, \citet[DeepChrome]{DeepChrome2016} visualized high-order combinatorial to make the model interpretable. Other examples of epigenetic information that can be utilized in gene expression prediction tasks include DNA methylation, miRNA, chromatin features, \textit{etc}. 

Generative models were also adopted due to the ability to capture high-order, latent correlations. For example, to explore hypothetical gene expression profiles under various types of molecular and genetic perturbation, \citet{way2017vaeCancer} trained a VAE on The Cancer Genome Atlas \cite[TCGA;][]{weinstein2013cancer} pan-cancer RNA-seq data to capture biologically-relevant features. They have another previous work that evaluates VAEs of different architectures, provided with comparison among VAEs, PCA, ICA, non-negative matrix factorization (NMF), and aforementioned ADGAE \citep{way2017evaluating}. \yx{Having the emergent capability to integrate long-range interactions in the genome, Generative Language Models such as \citet{enformer} also claimed to improve gene expression prediction accuracy from DNA sequences, leading to more accurate variant effect predictions on gene expression for both natural genetic variants and saturation mutagenesis measured by massively parallel reporter assays.}

\subsection{Regulatory Genomics}
Gene expression regulation is the cellular process that controls the expression level of gene products (RNA or protein) to be high or low. It increases the versatility of an organism so as to allow it to react towards and adapt to the surrounding environment.
The underlying interdependencies behind the sequences limit the flexibility of conventional methods, but deep networks that could model over-representation of sequence information have the potential to allow regulatory motifs to be identified according to their target sequences.

\subsubsection{Promoters and Enhancers}

The most efficient way of gene expression regulation for an organism is at the transcriptional level, which occurs at the early stage of gene regulation. Enhancers and promoters are two of the most well-characterized types of functional elements in the regions of non-coding DNA, which belong to cis-regulatory elements (CREs).
%
%
Readers can visit \citet{wasserman2004CREs1} and \citet{CREs22015identification} for a review of early approaches for the identification of CREs.

Promoters locate near the transcription start sites of genes and thereby initiate the transcription of particular genes. Conventional algorithms still perform poorly on promoter prediction, while the prediction is always accompanied by a high false positive rate \citep{fickett1997eukaryotic}. The compensation for sensitivity is usually achieved at the cost of specificity and renders the methods not accurate enough for applications. One initial work by \citet{horton1992assessment} applied neural networks to predict E. coli promoter sites and provided a comparison of neural networks versus statistical methods. \citet{CNNpromoters1996early} also applied neural networks to promoter recognition, albeit assisted with some rules which use the gene context information predicted by GRAIL. These early works of deep learning models were not noticeable enough to demonstrate a clear edge over the weight matrix matching methods. One recent study by \citet{CNNpromoters2017} used a CNN with no more than three layers well demonstrated the superiority of CNN over conventional methods in promoter recognition of five distant organism.
Their trained model has been implemented as a web application called CNNProm. 
A more latest CNN-based model for enhancer prediction applied transfer learning setting on different species/datasets \citep{cohn2018enhancer}. Another highlight of their work lies in the design of adversarial training data.

PEDLA was developed by \citet{liu2016pedla} as an algorithmic framework for enhancer prediction based on deep learning. It is able to directly learn from heterogeneous and class-imbalanced data an enhancer predictor that can be generalized across multiple cell types/tissues. The model has an embedded mechanism to handle class-imbalanced problems in which the prior probability of each class is directly approximated from the training data. 
PEDLA was first trained on 9 types of data in H1 cells, then further extended with an iterative scheme that manages to generalize the predictor across various cell types/tissues. PEDLA was also compared with and outperformed some of the most typical methods for predicting enhancers.

\citet[DeepEnhancer]{deepenhancer2016} adopted CNNs that surpass previous sequence-based SVM methods on the task of identifying enhancers from background genomic sequences. They compared different designs of CNNs and concluded the effectiveness of max-pooling and batch normalization for improving classification accuracy, while they also pointed out that simply increasing the depth of deep architectures is not useful if being inappropriately designed. Their final model has been fine-tuned on ENCODE cell type-specific enhancer datasets from the model trained on the FANTOM5 permissive enhancer dataset by applying transfer learning. 

\citet{yang2017biren} showed the possibility of predicting enhancers with DNA sequence alone with the presentation of BiRen, a hybrid of CNN and RNN. While demonstrating the possibility, there seems to be room to improve BiRen with the techniques that enable deep learning over heterogeneity data (\textit{e.g.} see Section~\ref{method:hetero}) since BiRen still exhibits weaker predictive performance in comparison to the methods that consider the cell-type-/tissue-specific enhancer markers explicitly. 

Deep Feature Selection (DFS) is an attempt taken by \citet{DFSi2015deep} to introduce sparsity to deep architectures. Conventionally, the sparseness is achieved by adding a regularization term (\textit{e.g.}, LASSO, Elastic Net). \citet{DFSi2015deep} took a novel approach by which they can automatically select an active subset of features at the input level to reduce the feature dimension.
This is implemented as an additional sparse one-to-one (point-wise product) linear layer between the input data and the input layer of the main model. 
DFS is widely applicable to different deep architectures. For example, \citet{DFSi2015deep} demonstrated MLP-based DFS (shallow DFS), DNNs based DFS (Deep DFS), and pointed out that when back-propagation does not perform well for deep networks, people can resort to stacked contractive autoencoder (ScA) and DBN based DFS models that pre-trained layer-wisely in a greedy way before fine-tuned by back-propagation. The author developed an open-source package of DFS and illustrated the superiority of DFS over Elastic Net and Random Forest in the identification of enhancers and promoters. \citet{CREs2016genome} further implemented a supervised deep learning package named DECRES, a feed-forward neural network based on DFS, for genome-wide detection of regulatory regions. 

Enhancer-promoter interaction predictions are always based on non-sequence features from functional genomic signals. \citet[SPEID]{singh2016EnhancerPromoter} proposed the first deep learning approach to infer enhancer-promoter interactions genome-wide from only sequence-based features, as well as the locations of putative enhancers and promoters in a specific cell type. Their model was demonstrated to be superior to DeepFinder, which is based on machine learning \citep{whalen2016enhancer}. This hybrid model consists of two parts. The first part accounts for the differences of underlying features that could be learned between enhancers and promoters and thus treats enhancers and promoters separately at input by two branches, where each branch is a one-layer CNN followed by a rectified linear unit (ReLU) activation layer. The second part is an LSTM that is responsible for identifying informative combinations of the extracted subsequence features. Their work provided insights into the long-range gene regulation determined from the sequences.

\yx{LLMs has made significant progress in promoter and enhancer-related prediction tasks. \citet{dnabert} utilized a BERT architecture to effectively predict proximal and core promoter regions and the successor of this work \cite{dnabert-2} claimed to achieve optimal performance in the Core Promoter Detection task. \citet{nucleotide-transformer} performed comprehensive benchmarking on 17 datasets including predicting regulatory elements for enhancers and promoters for several transformer models.}



\subsubsection{Splicing}

Splicing refers to the editing of pre-messenger RNA so as to produce a mature messenger RNA (mRNA) that can be translated into a protein.
This process effectively add up to the diversity of protein isoforms.
Predicting "splicing code" 
aims to understand how splicing regulates and manifests the functional changes of proteins, and is crucial for understanding different ways of how proteins are produced.

Initial machine learning attempts included naive Bayes model \citep{barash2010deciphering} and two-layer Bayesian neural network \citep{xiong2011bayesian} that utilized over a thousand sequence-based features. Early applications of neural networks in regulatory genomics simply replaced a classical machine learning approach with a deep model. For example, \citet{Xiong2015RNAsplicing} adopted a fully connected feed-forward neural network 
trained on exon-skipping events in the genome that can predict splicing regulation for any mRNA sequence. They applied their model to analyze more than half a million mRNA splicing codes for the human genome, 
and discovered many new disease-causing candidates while thousands of known disease-causing mutations were successfully identified. This is a case where high performance mainly results from a proper data source rather than a descriptive model design. \citet{rbm2015boosted} presented a DBN-based approach that is capable of dealing with class-imbalanced data to predict splice sites while also identifying non-Canonical splice sites. They also proposed a new training method called boosted contrastive divergence with categorical gradients and showed by their experiments its ability to improve prediction performance and shorten runtime compared to contrastive divergence or other methods. \cm{ \citet{gao2021} developed an approach based on CNN and uses sequence signatures to identify gene targets of a therapeutic for human splicing disorders.}

In many cases happens the phenomenon of alternative splicing. That is, a single gene might end up coding for multiple unique proteins by varying the exon composition of the same mRNA during the splicing process. This is a key post-transcriptional regulatory mechanism that affects gene expression and contributes to proteomic diversity \citep{juan2016mechanisms}. 
\citet{splicingCode2014} developed a DNN model containing three hidden layers to predict alternative splicing patterns in individual tissues, as well as across-tissue differences.
The hidden variables of the model are designed to include cellular context (tissue types) information to extract genomic features. This is one of the initial works that adapt deep learning for splicing prediction.
Work by \citet{jha2017integrative} based on previously developed BNN \citep{xiong2011bayesian} and DNN \citep{splicingCode2014} models to design an integrative deep learning model for alternative splicing. They viewed previous work as the baseline on their original dataset, and further developed these models by integrating additional types of experimental data (\textit{e.g.} tissue type) and proposed a new target function. Their models are able to identify splicing regulators and their putative targets, as well as infer the corresponding regulatory rules 
directly from the genomic sequence.



\subsubsection{Transcription Factors and RNA-binding Proteins}

Transcription factors (TFs) refer to proteins that bind to promoters and enhancers on DNA sequence, and RNA-binding proteins, as the name suggested, are both crucial regulatory elements in biological processes.
%
Current high-throughput sequencing techniques for selecting candidate binding targets for certain TFs are restricted by the low efficiency and high cost \cite{opportunitiesObstacles2017}.
Researchers seeking computational approaches for TF binding sites prediction on DNA sequences initially utilized consensus sequences or its alternative, position weight matrices \citep{stormo2000dna}. Later machine learning methods SVM using k-mer features \citep{gkmSVM2014}, \citep{setty2015seqgl} surpassed previous generative models.

Many existing deep learning methods approach TFBS prediction tasks through convolutional kernels. 
\citet[DeepBind]{DeepBind} have shown successful using CNN models in large-scale problems of TFBS tasks. 
\citet{chen2017predicting} combined the advantage of representation learning from CNN and the explicity of reproducing kernel Hilbert space to introduce the Convolutional Kernel Networks to predict TFBS with interpretability. 
\citet{cnnIntro2016} conducted a systematic analysis of CNN architectures for predicting DNA sequence binding sites based on large TF datasets. \citet{Dashboard} further explored CNNs, RNNs, and the combination of the two in the task of TFBS with comprehensive discussion and visualization techniques. Admittedly that CNNs can well capture most sequential and spatial features in DNA sequences, but recurrent networks as well as bidirectional recurrent networks are useful when accounting for motifs in both directions of the sequence. Motivated by the symmetry of double-stranded DNA, which means that identical patterns may appear on one DNA strand and its reverse complement, 
\citet{shrikumar2017reverse} proposed a traditional convolution-based model which shares parameters of forward and reverse-complement versions of the same DNA sequences, and have shown robust on in vivo TFBS prediction tasks using chromatin ChIP-seq data. This is a novel work that tailors conventional neural networks to consider motifs through bidirectional characterizations.

In addition to convolutional neural networks, which proved powerful as long as being appropriately designed according to the specific problem, some other approaches deal with different feature extraction or multiple data sources. Cross-source data usually share common knowledge at a higher abstraction level beyond the basic observation and thus need to be further integrated by the model. \citet{zhang2015deep} proposed a multi-modal deep belief network that is capable of automatic extraction of structural features from RNA sequences; they first successfully introduce tertiary structural features of RNA sequences to improve the prediction of RNA-binding proteins interaction sites. Another multi-modal deep learning model for the same purpose was developed by \citet[iDeep]{pan2017rna}. This model consists of DBNs and CNNs to integrate lower-level representations extracted from different data sources. \citet[gkm-DNN]{cao2017gkmDNN} based on gapped k-mers frequency vectors (gkm-fvs) to extract informative features. The gkm-fvs after normalization are taken as input for a multi-layer perceptron model trained by the standard error back-propagation algorithm and mini-batch stochastic gradient descent. By taking advantage of both gapped k-mer methods and deep learning, gkm-DNN achieved overall better performance compared with gkm-SVM. \citet[TFImpute]{qin2017imputation} proposed a CNN-based model that utilizes domain adaptation methods, which are discussed in more detail in Section \ref{sub-sec:transfer}, to predict TFs in new cell types by models trained unsupervisedly on TFs where ChIP-seq data are available. \yx{\citet{dnabert} and \citet{dnabert-2} fine-tuned BERT-based models to more accurately predict TFBSs, on both human and mouse genomic tracks.}

\subsection{Functional Genomics}

\subsubsection{Mutations and Functional Activities}

One of the shortcomings of previous approaches for predicting the functional activities from DNA sequences is the insufficient utilization of positional information. 
Though \citet{gkmSVM2014} upgraded the k-mer method by introducing an alternative gapped k-mers method (gkm-SVM), the improvement is not remarkable since the DNA sequence is still simply represented as vectors of k-mer counts without considering the position of each segment in the sequence. 
Though position-specific sequence kernels exist, they map the sequence into much higher dimension space and are thus not efficient enough \citep{Basset2016Kelley}. 

In contrast to conventional methods, deep learning methods such as CNNs naturally account for positional relationships between sequence signals and are computationally efficient. \citet[Basset]{Basset2016Kelley} presented an open-source CNN-based package trained on genomics data of 164 cell types, and remarkably improved the prediction for functional activities of DNA sequences. Basset enables researchers to perform the single sequencing assay and annotate mutations in the genome with present chromatin accessibility learned at the same time.
\citet[DeepSEA]{deepsea2015predicting} contributed another open-source deep convolutional network for predicting from only genomic sequence the functional roles of non-coding variants on histone modifications, TFBS, and DNA accessibility of sequences with high nucleotide resolution. \yx{Since CNN-based methods might require a certain amount of supervised training data, \citet{benegas2022dna} utilized pre-trained DNA language models to perform zero-shot non-coding variant effects prediction and the results outperformed previous approaches where vast amounts of functional genomics data are required for training.}

The effects of mutations are usually predicted by site-independent or pairwise models, but these approaches do not sufficiently model higher-order dependencies.
\citet[DeepSequence]{riesselman2017deepGenerative} took a generative approach to tract mutation effects that are beyond pairwise by biologically-motivated Beyasian deep latent networks. They introduced latent variables on which DNA depends and visualized model parameters to illustrate the structural proximity and amino acid correlations captured by DeepSequence.




%





\subsubsection{Subcellular Localization}
Subcellular localization is to predict the subcellular compartment in a protein that resides in the cell from its biological sequence. In order to interact with each other, proteins need to at least temporarily inhabit physically adjacent compartments, therefore, the knowledge of protein location sheds light on where a protein might function as well as what other proteins it might interact with \citep{sherLoc2007}. Most previous methods rely on SVMs and involve hand-generated features. For example, \citet[SherLoc]{sherLoc2007} integrated different sequence and text-based features, and \citet[BaCelLo]{pierleoni2006bacello} developed a hierarchy of binary SVMs. \citet{meinken2012computational} reported on previous tools and \citet{wan2015machine} covered the machine learning approaches for subcellular localization.
M
Some early deep learning works have shifted from SVMs to neural networks, such as \citet{emanuelsson2000predicting} and  \citet{hawkins2006detecting}. \citet{mooney2011sclpred} based on an N-to-1 neural network to develop a subcellular localization predictor (SCLpred). 
\citet{sonderby2015convolutionalLSTM} adopted LSTM to predict protein subcellular locations from only sequence information with high accuracy. 
They further enhanced the model by adding convolutional filters before LSTM as a motif extractor and introducing the attention mechanism that forces the LSTM to focus on particular segments of the protein. The validity of their convolutional filters and attention mechanisms were visualized in experiments. \citet{almagro2017deeploc} proposed a similar integrative hybrid model DeepLoc consisting of four modules, including CNN, BLSTM, attention scheme and q fully connected dense layer. \cm{\citet{cite-key} utilized vector quantized VAE architecture to encode high-resolution features of protein subcellular localization without the need for prior knowledge, categories or annotations.}

High-throughput microscopy images are a rich source of biological data that remains to be better exploited. One of the important utilization of microscopy images is the automatic detection of the cellular compartment. \citet[DeepYeast]{parnamaa2017Subcellular} devised an eleven-layer deep model for fluorescent protein subcellular localization classification in yeast cells, of which eight convolutional layers are succeeded by three fully connected layers. Internal outputs of the model are visualized and interpreted from the perspective of image characteristics. The author concluded that the low-level network functions as a basic image feature extractor, 
while higher layers account for separating localization classes.

%
%

\subsection{Structural Genomics}

\subsubsection{Structural Classification of Proteins}
Proteins usually share structural similarities with other proteins, some of which have a common evolutionary origin \citep{lo2000scop}. 
Classification of protein structure can be traced back to the 1970s aiming to comprehend the process of protein folding and protein structure evolution \citep{andreeva2010structural}. 
Grouping proteins into structural or functional categories also facilitates the understanding of an increasing number of the newly sequenced genome.

Early methods for similarity measures 
mostly rely on sequence properties (\textit{i.e.} alignment-based), such as FASTA \citep{FASTA1988improved}, BLAST \citep{BLASTaltschul1990}, or PSI-BLAST \citep{altschul1997psiBLAST}, and were then upgraded by leveraging profiles derived from multiple sequence alignments and position-specific scoring matrices (PSSM) in addition to raw sequences \citep{rangwala2005profile}, or discriminative models like SVM \citep{liao2003combining}. For example,  \citet{cang2015topological} adopted SVM with a topological approach utilizing persistent homology to extract features for the classification of protein domains and superfamilies. Other top-performing deep learning works also rely on protein homology detection (one can visit \citet{chen2016comprehensive} for a review) to deduce the 3D structure or function of a protein from its amino acid sequence. \citet{proFam2007hochreiter} suggested a model-based approach that uses LSTM for homology detection. 
Their model makes similarity measures such as BLOSUM or PAM matrices, not a priori fixed, but instead suitably learned by LSTM with regard to each specific classification task. \citet[ProDec-BLSTM]{liu2017protein} conducted a similar work on protein remote homology detection, and showed an improvement using BLSTM instead of LSTM \citep{proFam2007hochreiter}. One drawback of homology-based approaches for fold recognition is the lack of a direct relationship between the protein sequence and the fold since current methods substantially rely on the fold of known template protein to classify the fold of new proteins \citep{hou2017deepsf}. Therefore, \citet[DeepSF]{hou2017deepsf} proposed a deep 1D CNN for fold classification directly from protein sequences.

There are also some works based on available gene function annotation vocabularies (\textit{e.g.} Gene Ontology \citep{park2005protein}) to perform protein classification \citep{ashburner2000geneOntology}. By Similar motivation, BioVec \citep{skipgram2015continuous} was designed as a deep learning method to compute a distributed representation of biological sequences with general genomic applications such as protein family classification. Each sequence is embedded in a high-dimension vector by BioVec, then the classification of protein families can reduce to a simple classification task.


\subsubsection{Protein Secondary Structure}
\lx{Protein Secondary Structure (SS) refers to the local spatial structure formed by interaction between nearby stretches of a polypeptide chain. The protein SS encodes information for predicting biophysical properties of amino acid residues, higher-level protein structure (e.g., tertiary structure), protein function and evolution. It is traditionally described by either a 3-state model \citep{3state1951pauling}, or an 8-state model by DSSP algorithm \citep{8state1983kabsch}. The former labels each residue to be in one of three states: Helix, Strand, or Coil; while the 8-state model expands to eight different states for a more fine-grained description of the spatial environment and chemical bonding of each amino acid. Q3 and Q8 accuracies are the widely adopted metrics to evaluate any model performance, which represents the percentage of correctly predicted secondary conformation of amino acid residues. An alternative measure for 3-state prediction is the segment of overlap (SOV) score \citep{SOV1999zemla}. The reasonable goal of SS prediction is suggested by \citet{rost1994redefining} as a Q3 accuracy above 85\%}

Before deep learning became popular for protein SS prediction, machine learning approaches including probabilistic graphical models \citep{bayesSS2000schmidler,hiddenSS2011,graphicalSS2004}, hidden
Markov models \citep{hiddenSS2011} and SVMs \citep{SVMss2001novel,SVMss20032,SVMss2003} were widely adopted. 
At that nascent age of neural networks, one of the earliest applications 
developed a shallow feed-forward network that predicts protein SS and homology from the amino acid sequences \citep{1988earliestProteinSS}. 
Other works for SS prediction adopted similar or slightly enhanced neural networks \citep{basicNN11989holley,basicNN31990kneller}. \citet{basicNN21988} conducted one of the influential works for 3-state prediction, reaching a Q3 accuracy of 64.3\%. They based on the fully connected neural networks to develop a cascaded architecture, taking as input window DNA sequences with orthogonal encoding.
There was no significant progress for 3-state prediction accuracy by neural networks until being improved to 70.8\% by \citet{70proteinSS1993,rost1993improved}.
Claimed of the marginal influence of free parameters in the model, \citet{70proteinSS1993} accredited their improvement to leveraging evolutionary information encoded in the input profiles derived from multiple alignments.  \citet{riis1996structured} achieved a practically identical performance by a structured neural network. They designed specific networks for each SS class according to biological knowledge and the output prediction was made from filtering and ensemble averaging. Based on the PSSM generated by PSI-BLAST, \citet[PSIPRED]{PSIPRED1999jones} used a 2-stage neural network to obtain an average Q3 score of around 77\%. Other popular deep learning methods such as bidirectional RNNs were also widely applied for protein SS prediction \citep{basicNN41999exploiting,rnnSS2002pollastri,rnnSS2014magnan}.

Emergent deep architectures for protein SS prediction have been widely explored 
with more prior knowledge and various features available. 
\citet[SPINE X]{faraggi2012spine} proposed an iterative six-step model, of which the neural network of each step follows a similar structure and is designed for each specific purpose. 
\citet{2015spencer} trained a deep belief network model, in which an additional hidden layer is constructed to facilitate the unsupervised layer-by-layer initialization of the Restricted Boltzmann Machine (RBM). \citet{cascadSS2016} designed a cascaded model, which leverages CNN to extract multi-scale local contextual features by different kernel sizes, then added a BRNN accounting for long-range dependencies in amino acid sequences to capture global contextual features. 

\citet[DeepCNF]{DeepCNF2016Wang} took a large step in improving Q3 accuracy above 80\% by extending conditional neural fields (CDFs) to include convolutional designs. DeepCNF is able to capture both sequence-structure relationships and protein SS label correlation among adjacent residues. They also achieved Q8 accuracy of around 72\%, outperforming Q8 accuracy of 66.4\% obtained by a supervised generative stochastic network \citep{66Q8ss2014}. 
\citet{busia2016protein} explored the model performance of 8-stated prediction from simple feed-forward networks to the adaptation of recent CNN architectures (\textit{e.g.} Inception, ReSNet, and DenseNet). They modified the convolution operators of different scales and residual connections of successful CNN models in computer vision to suit the protein SS prediction task and also highlighted the differences compared to vision tasks. As opposed to the above-mentioned DeepCNF \citep{DeepCNF2016Wang} that included interdependencies between labels of adjacent residues by Conditional Random Field (CRF), \citet{busia2016protein} conditions the current prediction on previously predicted labels by sequence-to-sequence modeling.

\wwp{A new class of Protein Language Model (ProtLM) has been proposed in recent years to utilize the power of large-scale, transformer-based language models. \citet{Rives2019BiologicalSA} introduced ESM-1b model, a BERT-like model trained on up to 250 million protein sequence data from UniRef50 and UniRef100 datasets \citep{Suzek2014UniRefCA} using the common masked language model (MLM) objective, and achieved 70\%+ Q8 accuracies at family, superfamily, and fold levels on a constructed test set derived from SCOPe database \citep{Fox2013SCOPeSC}. The ProtTrans model family \cite{Elnaggar2020ProtTransTC} introduced a series of transformer model architectures (Transformer-XL \citep{Dai2019TransformerXLAL}, BERT \citep{bert}, and T5 \citep{Raffel2019ExploringTL}) pretrained on protein sequences from UniRef \citep{Suzek2014UniRefCA} and Big Fantastic Database \citep{Steinegger2018ProteinlevelAI_BFD}; they also showed the potential of ProtLM completely independent of multiple sequence alignments (MSAs) features; their best-performing model achieved a Q3 accuracy of 74.1\% and Q8 accuracy of 60.7\% on CASP14 \citep{kryshtafovych2021critical}. ESM-2 model family \citep{lin2023evolutionary} was later introduce as an upgrade of the ESM-1b model in size (up to 15 billion parameters) and pushed the Q3 accuracy to 76.8\% and Q8 accuracy to 61.7\% on CASP14. The latest effort in scaling up ProtLM has resulted in a 100-billion-parameter model, xTrimoPGLM \citep{Chen2023xTrimoPGLMU1}, with a model backbone of General Language Model (GLM) \citep{Du2021GLMGL}; however, the authors also pointed out that the logarithmic increase of model performance on size has already shown saturation on specific task such as the Q3 SS prediction.}

\subsubsection{Contact Map}
\label{sec:contact_map}
A protein contact map is a binary 2D matrix denoting the spatial closeness of any two residues in the folded 3D protein structure. Predicting residue-residue contact is thus crucial to protein structure prediction, and has been early studied by shallow neural networks \citep{torracinta2016training}. 
Recent works proceeded to deeper networks. 
\citet{DSTnn2012deep} stacked together multiple standard three-layer feedforward networks sharing the same topology, taking into consideration both spatial and temporal features to predict protein residue–residue contact. \citet{ultraDeep2017} also developed an ultra-deep model to predict protein contacts from amino acids sequence. Their model consists of two deep residual neural networks that process 1D and 2D features separately and subsequently in order to consider both sequential and pairwise features in the whole model. 
\citet{zhang2017hicplus} and \citet{schreiber2017nucleotide} both contributed an open-source multi-modal CNN model for Hi-C contact map prediction. \citet[HiCPlus]{zhang2017hicplus} first interpolated the low-resolution Hi-C matrix to the size of the high, then trained their model to predict high- from the low-resolution matrix. The final output was recombined into the entire Hi-C interaction matrix.
\citet[Rambutan]{schreiber2017nucleotide} predicted Hi-C contacts at high resolution (1 kb) from nucleotide sequences and DNaseI assay signal data. Their model consists of two arms, with each arm processing one type of data independently. The learned feature maps are then concatenated for further combination with genomic distance in the dense layers. \citet{adhikari2018dncon2} proposed a two-layer CNN network that consumes PSSM based features as well as coevolutionary contact features to classify residue-residue contact into five distance bins (6-10 \r{A}).

\lx{
In recent years, ResNet \citep{ResNet2016deep} has been extensively used in contact map prediction. \citet{yang2020improvedtrrossetta} proposed trRossetta (based on Rossetta3 software \cite{leaver2011rosetta3}), which utilized known MSAs to guide the training of a 60-layer ResNet to classify inter-residue distance as well as inter-residue orientational angles into discrete value bins. DeepDist \citep{wu2021deepdist} showcased the potential of ResNet for direct prediction of residue-residue contact distance by combining outputs from four ResNet networks, each on one type of sequential or co-evolutional features from the input, and was trained successfully with a regression objective that minimizes the Mean-Squared Error (MSE) of predicted contact distance in absolute value.
}

\subsubsection{Protein Tertiary Structure and Quality Assessment}

The prediction of protein tertiary structure has proven crucial to human understanding of protein functions \citep{breda2007protein} and can be applied to, for instance, drug designs \citep{jacobson2004comparative}. However, experimental methods for determining protein structures, such as X-ray crystallography, are costly and sometimes impractical. Though the number of experimentally solved protein structures included in the protein data bank 
(PDB)(\url{https://www.rcsb.org/})
keeps growing, it only accounts for a small proportion of currently sequenced proteins \citep{kryshtafovych2009protein}. Thus, a potentially practical approach to fill 
the gap between the number of known protein sequences and the number of found protein structures is through computational modeling.

Two essential challenges in protein structure prediction include the sampling and the ranking of protein structural models \citep{cao2015large}. Quality assessment (QA) is to predict the absolute or relative quality of the protein models before the native structure is available so as to rank them. Some previous research, such as \citep[ProQ2]{proQ2-2012}
and \citep[ProQ3]{ProQ3-2016}, was conducted based on machine learning models. Recent deep learning-based work from \citet[ProQ3D]{uziela2017proq3d} achieved substantial improvement by replacing the SVMs in previous work with DNNs. As opposed to these existing methods that rely on energy or scoring functions, \citet{nguyen2014dl} based solely on geometry to propose a sparse stacked autoencoder classifier that utilizes the contact map. 
Another research by \citet{cao2016deepqa} adopted a deep belief network protein structure prediction. Their model could be used to evaluate the quality of any protein decoy. 
Local quality assessment remains to be substantially improved compared with global prediction \citep{shin2017prediction}. \citet{liu2016benchmarking} introduced three models based on SDAs as a benchmark of deep learning methods for assessing the quality of individual protein models.

\lx{Modern tertiary structure prediction systems typically pipeline functional modules that (i) query MSA for target protein sequence; (ii) predict contact map or residue-residue distance, and (iii) reconstruct 3-D structure based on predicted contact map under energy and physical constraints. The quality of these MSA-based systems can depend sensitively on the performance of the involved contact map prediction models (Section \ref{sec:contact_map}). MULTICOM \citep{hou2019proteinmulticom} extended the use of DNCON2 as their contact map prediction module by incorporating 1-D structural features, such as residue-level secondary structure labels, sequential features, and coevolutionary features; the system was upgraded in 2022 to MULTICOM2 \citep{liu2022improvingmulticom2} as the authors incorporated more deep-learning based modules, including using DeepDist in place of DNCON2 for contact map prediction, and achieved high system ranking (seventh out of 146 systems) in tertiary structure prediction in CASP14. ThreadAI \citep{zhang2020templatethreadai} also improved upon MULTICOM by adopting trRossetta instead of DNCON2 in their contact map prediction.

The AlphaFold \cite{Senior2020AlphaFold} and AlphaFold2 models \citep{jumper2021alphafold2} revolutionized the practicality of DNNs in predicting protein structure at atomic resolution: the authors constructed a novel two-stage DNN relying heavily the attention mechanism to predict directly the 3D coordinates of all heavy atoms in a given protein: the first stage encodes and combines both MSA and residue pairs features through a series of transformer-like blocks with attention module; the second stage module then builds upon the learned representations to refine a hypothesized 3D structure subjecting to evolutionary, physical, and geometrical constraints. As of the writing of this review, AlphaFold2 remains the best model for protein structure prediction on CASP14. Research since has revealed several limitations of the AlphaFold2 model: its performance coud suffer from predicting intrinsically disordered proteins \citet{ruff2021alphafolddisorder}; the performance on loop prediction is only high for short loops \citet{stevens2022alphafoldloop}; and significant degradation in performance was discovered on target sequence with few homologous counterparts in existing databases or when the MSAs are of low depth \citet{wang2022contactlowmsa}, although this problem was mitigated by MSA-Augmenter \citet{zhang2023enhancingmsaagumenter}, which is a transformer model trained on known MSA sequences to generate artificial MSA sequences that are used to augment training data used by AlphaFold2.

A more serious limitation with AlphaFold2 is the inability of predicting novel structures due to its dependence on known MSA features. ProtLM models which are independent of MSA features overcome this limitation trivially and has shown great potential in novel structure prediction. \citet{lin2023evolutionary} introduced ESMFold as an extension of ESM-2 model family with an added folding structure module; ESMFold depends only on embeddings learned through ESM-2 and showed decent performance (80\% of AlphaFold2) on CASP14. More competitive performance is achieved by newer ProtLM such as OmegaFold \citep{wu2022omegafold} and EMBER3D \citep{weissenow2022ember3d}. These models also showed much better inference time than AlphaFold2, with OmegaFold attaining sub-second prediction on proteins with sequence up to 1000 residues.
}

\section{Challenges and Opportunities}
\label{sec:opportunities}

With the discussion of the successes of applications of deep learning in genomics, now we proceed to discuss some current challenges. As deep learning models are usually over-parametrized, the performance can be conditional if the models are not appropriately designed according to the problem. There are multiple worthwhile considerations and techniques involving model architectures, feature extraction, data limitation, \textit{etc.}, which help deep learning models to better approach genomics. 
Here we briefly discuss some current challenges that deserve attention and several potential research directions that might shed on light the future development of deep learning applications in genomic research.

\subsection{The Nature of Data}

An inevitable challenge of transferring the success of deep learning in conventional vision or text data into genomics is raised due to the nature of the genomic data, such as the unavailability of true labels due to the lack of knowledge of the genetic process, the imbalanced case and control samples due to the rarity of a certain disease, and the heterogeneity of data due to the expensiveness of large-scale data collection.

\subsubsection{Class-Imbalanced Data}\label{class-imbal}
Large-scale biological data gathered from assorted sources are usually inherently 
class-imbalanced. Take epigenetic datasets for example, there are in nature much fewer DNA methylated regions (DMR) sites than non-DMR sites \citep{haque2014imbalanced}. It is also common in enhancer prediction problems where the number of non-enhancer classes overwhelmingly exceeds that of enhancer classes \citep{firpi2010discover,kleftogiannis2014deep}.
\hx{Methods that directly over-samples minor classes or under-samples majority classes have been attempted and proven successful. \citet{upsample2023} found that multiple machine learning models (SVM, MLP, Random Forest) benifits from Synthetic Minority Over-sampling Technique (SMOTE) in finding single nucleotide polymorphisms (SNPs).} This data-imbalance issue has also been encountered in machine learning methods \citep{yoon2005unsupervised,he2013imbalanced}, while ensemble methods appear to be powerful \citep{haque2014imbalanced}. \citet{sun2013image} applied the undersampling method together with a majority vote to address the imbalanced data distribution inherent in gene expression 
image annotation tasks.
In deep learning approaches, \citet{al2016transfer} based on boosting to propose an instance-transfer model to reduce the class-imbalanced influence while also improving the performance by leveraging data from an auxiliary domain. \hx{Combining conventional DNNs with a Deep Decision Tree classifier, \citet{hdnn2021} proposes a Hybrid DNN architecture which addresses class-imbalance of certain RNA sequences by forcing the network to account for minor classes in the decision tree's hierarchical if-else cases.} In addition to resorting to ensemble approaches, 
researchers can manage to resolve class-imbalanced problems through model parameters or training processes. For instance, \citet[PEDLA]{liu2016pedla} used an embedded mechanism utilizing the prior probability of each class directly estimated from the training data to compensate for the imbalance of classes. \citet{rbm2015boosted} presented a method called boosted contrastive divergence with categorical gradients for training RBMs for class imbalanced prediction of splice junctions. \citet{singh2016EnhancerPromoter} performed data augmentation by slightly shifting each positive promoter or enhancer within the window since the true label is not sensitive to these minimal changes. They also designed the training procedure accordingly to avoid the high false positive rate resulting from the augmented dataset.


\subsubsection{Various Data Types}
Intuitively, integrating diverse types of data as discriminating features will lead to more predictive power of the models.
For example, \citet[PEDLA]{liu2016pedla} trained their model on nine types of data to identify enhancers, including chromatin accessibility (DNase-sseq), TFs and cofactors (ChIP-seq), histone modifications (ChIP-seq), transcription (RNA-Seq), DNA methylation (RRBS), sequence signatures, evolutionary conservation, CpG islands, and occupancy of TFBSs, resulting in better model performance in terms of multiple metrics compared with existing popular methods. \citet[DeepCpG]{DeepCpG2017Angermueller} predicted single-cell DNA methylation states 
by two disparate sub-networks designed accordingly for CpG sites and DNA sequences.

It pays off to manage to utilize the data of multiple views; though merging the information from various data sources challenges the models that could well integrate them, this effort might provide more information with a great chance. 
\hx{A review of data representations in genomics, transcriptomics, proteomics, metabolomics and epigenomics for computer scientists can be found in \citet{tsimenidis_omics_2022}, in which the format, type, and encoding of data from these disciplines are presented, together with their common feature extraction techniques.}
For more discussions on encompassing diverse data sources, we refer to multi-view learning in Section~\ref{sub-sec:multi-modal}.

\subsubsection{Heterogeneity and Confounding Correlations}
\label{method:hetero}
The data in most genomic applications involving medical or clinical are heterogeneous due to population subgroups, or regional environments. One of the problems of integrating these different types of data is the underlying interdependencies among these heterogeneous data. Covariates are sometimes confounding, and render the model prediction inaccurate.

The Genome-Wide Association Study (GWAS) is an example where both population-based confounders (population subgroups with different ancestry) and individual relatednesses produce spurious correlations among SNPs to the trait of interest. 
Most existing statistical methods estimate confounders before performing causal inference. These methods are based on linear regression \citep{yu2006unified,astle2009population}, linear mixed model (LMM) \citep{kang2010variance,yang2014advantages}, or others \citep{song2015testing}. \citet{wang2017variable} tried to upgrade LMM and tested it on biological variable selection and prediction tasks. Though these LMM-based models \citep[\textit{e.g.} FaST-LMM,][]{lippert2011fast} are favored by some researchers and mathematically sufficient, their power pales when facing multiple nonlinear confounding correlations. The assumed Gaussian noise might overshadow true underlying causals, and LMM also fails to literally model the variable correlations. A seemingly more reliable approach is to through generative modeling, \textit{e.g.} \citet{hao2015probabilistic}. \citet{tran2017implicit} and \citet{louizos2017causal} are all based on variational inference to present an implicit causal model for encoding complex, nonlinear causal relationships, with consideration of latent confounders. \citet{tran2017implicit} optimized their model iteratively to estimate confounders and SNPs, and their simulation study suggested a significant improvement.

From the methodology perspective, several deep learning methods that are not designed exclusively for confounder correction, such as the domain adversarial learning \citep{ganin2016domain}, select-additive learning \citep{wang2017select}, and confounder filtering \citep{wu2018fair}, can be re-used, once the identification of confounder is presented. 


\subsection{Feature Extraction}

Deep learning that performs automatic feature extraction saves great efforts of choosing hand-engineered features, 
\citet{3dCNN2017Torng} also discussed the superiority of automatically generated features over manually selected features. However, in practice, it is unfortunately time-consuming to directly learn features from genomic sequences when complex interdependences and long-range interactions are taken into consideration. Researchers might still resort to task-specific feature extraction before automatic feature detection, 
which could strongly facilitate the model if skillfully designed. 

\subsubsection{Mathematical Feature Extraction}
%
%
Techniques borrowed from mathematics have great potential to interpret the complex biological structures behind data that otherwise will hinder the generalization of deep learning. For example, topology is a promising choice to untangle the geometric complexity underlying the 3D biomolecular structure of protein \citep{zixuanTopo2017}, and homology detection has been widely applied to protein classification problems \citep{proFam2007hochreiter,cang2015topological}. DeepMethyl \citep{TopoCpG2016} was developed as deep learning software using features derived from 3D genome topology and DNA sequence patterns. It is based on SDAs and is applied to predict methylation states of DNA CpG dinucleotides. \citet{zixuanTopo2017} introduced element-specific persistent homology (ESPH) into CNNs to predict protein-ligand binding affinities and protein stability changes upon mutation, including globular protein mutation impacts and membrane protein mutations impact. \hx{Finally, in making feature extraction techniques from popular literatures generally available to the genomics research community, \citet{Bonidia2023} made available a novel software package \textit{MathFeature}, which implements 20 mathematical descriptors and 17 convential descriptors, used to numerically encode long gene sequences.}

\subsubsection{Feature Representation}
By the conceptual analogy of the fact that humans communicate through languages, biological organisms convey information within and between cells through information encoded in biological sequences. To understand this language of life, \citet{skipgram2015continuous} designed BioVec, an unsupervised data-driven feature representation method, which embeds each trigram of biological sequence in a 100-dimensional vector that characterizes biophysical and biochemical properties of sequences. 
BioVec was trained by a variant of MLP adapted from word2vec \citep{mikolov2013distributed,mikolov2013efficient}, a typical method in natural language processing.
\citet{ng2017dna2vec} further utilized shallow two-layer neural networks to compute the representation of variable-length k-mers of DNA sequences that is consistent across different lengths. 
In contrast to representation by BioVec for individual kmers, \citet{seq2vec2016kimothi} based on doc2vec algorithm, an extension of word2vec, to proposed distributed representation of complete protein sequence, and successfully applied to protein classification following the settings of \citet{skipgram2015continuous}. \hx{Another feature representation technique was proposed by \citet{Hao_Jing_Sun_2023} on cancer survival prediction from genetic sequences. The authors specifically represented the shared and unique features from DNA, mRNA and miRNA, and leveraged the consistent and complement information in these features to improve prediction accuracy.} These types of feature representation have the potential to facilitate the work of genomics.

\section{Conclusion and Outlook}
\label{sec:conclusion}

Genomics is a challenging application area of deep learning that encounters 
unique challenges compared to other fields such as vision, audio, and text processing, since we have limited abilities to interpret genomic information but expect from deep learning a superhuman intelligence that explores beyond our knowledge. Yet deep learning is undoubtedly an auspicious direction that has constantly rejuvenated and moved forward genomic research in recent years. As discussed in this review, recent breakthroughs of deep learning applications in genomics have surpassed many previous state-of-the-art computational methods with regard to predictive performance, though slightly lag behind traditional statistical inferences in terms of interpretation.

Current applications, however, have not brought about a watershed revolution in genomic research. The predictive performances in most problems have not reached the expectation for real-world applications, and neither have the interpretations of these abstruse models elucidate insightful knowledge. A plethora of new deep learning methods are constantly being proposed but await artful applications in genomics. By careful selection of data sources and features or appropriate design of model structures, deep learning can be driven towards a bright direction which produces a more accurate and interpretable prediction. We need to bear in mind numerous challenges beyond simply improving predictive accuracy to seek essential advancements and revolutions in deep learning for genomics.

\section*{Acknowledgement}
\acks{
We would like to acknowledge the inspiration from a course instructed by \cite{cs273b}, and two reviews contributed by \cite{min2017deep} and \cite{opportunitiesObstacles2017}. A collaboratively written review paper on deep learning, genomics, and precision medicine, now available at \url{https://greenelab.github.io/deep-review/}
}. 




\newpage
\label{app:theorem}



\vskip 0.2in
\bibliography{ref}

\begin{thebibliography}{288}
\providecommand{\natexlab}[1]{#1}
\providecommand{\url}[1]{\texttt{#1}}
\expandafter\ifx\csname urlstyle\endcsname\relax
  \providecommand{\doi}[1]{doi: #1}\else
  \providecommand{\doi}{doi: \begingroup \urlstyle{rm}\Url}\fi

\bibitem[Adhikari et~al.(2018)Adhikari, Hou, and Cheng]{adhikari2018dncon2}
Badri Adhikari, Jie Hou, and Jianlin Cheng.
\newblock Dncon2: improved protein contact prediction using two-level deep convolutional neural networks.
\newblock \emph{Bioinformatics}, 34\penalty0 (9):\penalty0 1466--1472, 2018.

\bibitem[Al-Stouhi and Reddy(2016)]{al2016transfer}
Samir Al-Stouhi and Chandan~K Reddy.
\newblock Transfer learning for class imbalance problems with inadequate data.
\newblock \emph{Knowledge and information systems}, 48\penalty0 (1):\penalty0 201--228, 2016.

\bibitem[Alipanahi et~al.(2015)Alipanahi, Delong, Weirauch, and Frey]{DeepBind}
Babak Alipanahi, Andrew Delong, Matthew~T. Weirauch, and Brendan~J. Frey.
\newblock Predicting the sequence specificities of dna- and rna-binding proteins by deep learning.
\newblock \emph{Nat Biotech}, 33\penalty0 (8):\penalty0 831--838, Aug 2015.
\newblock ISSN 1087-0156.
\newblock URL \url{http://dx.doi.org/10.1038/nbt.3300}.
\newblock Computational Biology.

\bibitem[Almagro~Armenteros et~al.(2017)Almagro~Armenteros, S{\o}nderby, S{\o}nderby, Nielsen, and Winther]{almagro2017deeploc}
Jose~Juan Almagro~Armenteros, Casper~Kaae S{\o}nderby, S{\o}ren~Kaae S{\o}nderby, Henrik Nielsen, and Ole Winther.
\newblock Deeploc: prediction of protein subcellular localization using deep learning.
\newblock \emph{Bioinformatics}, 33\penalty0 (21):\penalty0 3387--3395, 2017.

\bibitem[Altschul et~al.(1990)Altschul, Gish, Miller, Myers, and Lipman]{BLASTaltschul1990}
Stephen~F Altschul, Warren Gish, Webb Miller, Eugene~W Myers, and David~J Lipman.
\newblock Basic local alignment search tool.
\newblock \emph{Journal of molecular biology}, 215\penalty0 (3):\penalty0 403--410, 1990.

\bibitem[Altschul et~al.(1997)Altschul, Madden, Sch{\"a}ffer, Zhang, Zhang, Miller, and Lipman]{altschul1997psiBLAST}
Stephen~F Altschul, Thomas~L Madden, Alejandro~A Sch{\"a}ffer, Jinghui Zhang, Zheng Zhang, Webb Miller, and David~J Lipman.
\newblock Gapped blast and psi-blast: a new generation of protein database search programs.
\newblock \emph{Nucleic acids research}, 25\penalty0 (17):\penalty0 3389--3402, 1997.

\bibitem[Andreeva and Murzin(2010)]{andreeva2010structural}
Antonina Andreeva and Alexey~G Murzin.
\newblock Structural classification of proteins and structural genomics: new insights into protein folding and evolution.
\newblock \emph{Acta Crystallographica Section F: Structural Biology and Crystallization Communications}, 66\penalty0 (10):\penalty0 1190--1197, 2010.

\bibitem[Angermueller et~al.(2017)Angermueller, Lee, Reik, and Stegle]{DeepCpG2017Angermueller}
Christof Angermueller, Heather~J. Lee, Wolf Reik, and Oliver Stegle.
\newblock Deepcpg: accurate prediction of single-cell dna methylation states using deep learning.
\newblock \emph{Genome Biology}, 18\penalty0 (1):\penalty0 67, Apr 2017.
\newblock ISSN 1474-760X.
\newblock \doi{10.1186/s13059-017-1189-z}.
\newblock URL \url{https://doi.org/10.1186/s13059-017-1189-z}.

\bibitem[Asgari and Mofrad(2015)]{skipgram2015continuous}
Ehsaneddin Asgari and Mohammad~RK Mofrad.
\newblock Continuous distributed representation of biological sequences for deep proteomics and genomics.
\newblock \emph{PloS one}, 10\penalty0 (11):\penalty0 e0141287, 2015.

\bibitem[Ashburner et~al.(2000)Ashburner, Ball, Blake, Botstein, Butler, Cherry, Davis, Dolinski, Dwight, Eppig, et~al.]{ashburner2000geneOntology}
Michael Ashburner, Catherine~A Ball, Judith~A Blake, David Botstein, Heather Butler, J~Michael Cherry, Allan~P Davis, Kara Dolinski, Selina~S Dwight, Janan~T Eppig, et~al.
\newblock Gene ontology: tool for the unification of biology.
\newblock \emph{Nature genetics}, 25\penalty0 (1):\penalty0 25--29, 2000.

\bibitem[Astle et~al.(2009)Astle, Balding, et~al.]{astle2009population}
William Astle, David~J Balding, et~al.
\newblock Population structure and cryptic relatedness in genetic association studies.
\newblock \emph{Statistical Science}, 24\penalty0 (4):\penalty0 451--471, 2009.

\bibitem[Avsec et~al.(2021)Avsec, Agarwal, Visentin, Ledsam, Grabska-Barwinska, Taylor, Assael, Jumper, Kohli, and Kelley]{enformer}
{\v Z}iga Avsec, Vikram Agarwal, Daniel Visentin, Joseph~R. Ledsam, Agnieszka Grabska-Barwinska, Kyle~R. Taylor, Yannis Assael, John Jumper, Pushmeet Kohli, and David~R. Kelley.
\newblock Effective gene expression prediction from sequence by integrating long-range interactions.
\newblock \emph{bioRxiv}, 2021.
\newblock \doi{10.1101/2021.04.07.438649}.
\newblock URL \url{https://www.biorxiv.org/content/early/2021/04/08/2021.04.07.438649}.

\bibitem[Bae et~al.(2015)Bae, Jayaraman, and Walsh]{bae2015genetic}
Byoung-Il Bae, Divya Jayaraman, and Christopher~A Walsh.
\newblock Genetic changes shaping the human brain.
\newblock \emph{Developmental cell}, 32\penalty0 (4):\penalty0 423--434, 2015.

\bibitem[Baldi et~al.(1999)Baldi, Brunak, Frasconi, Soda, and Pollastri]{basicNN41999exploiting}
Pierre Baldi, S{\o}ren Brunak, Paolo Frasconi, Giovanni Soda, and Gianluca Pollastri.
\newblock Exploiting the past and the future in protein secondary structure prediction.
\newblock \emph{Bioinformatics}, 15\penalty0 (11):\penalty0 937--946, 1999.

\bibitem[Barash et~al.(2010)Barash, Calarco, Gao, Pan, Wang, Shai, Blencowe, and Frey]{barash2010deciphering}
Yoseph Barash, John~A Calarco, Weijun Gao, Qun Pan, Xinchen Wang, Ofer Shai, Benjamin~J Blencowe, and Brendan~J Frey.
\newblock Deciphering the splicing code.
\newblock \emph{Nature}, 465\penalty0 (7294):\penalty0 53--59, 2010.

\bibitem[Barretina et~al.(2012)Barretina, Caponigro, Stransky, Venkatesan, Margolin, Kim, Wilson, Leh{\'a}r, Kryukov, Sonkin, et~al.]{barretina2012cancer}
Jordi Barretina, Giordano Caponigro, Nicolas Stransky, Kavitha Venkatesan, Adam~A Margolin, Sungjoon Kim, Christopher~J Wilson, Joseph Leh{\'a}r, Gregory~V Kryukov, Dmitriy Sonkin, et~al.
\newblock The cancer cell line encyclopedia enables predictive modelling of anticancer drug sensitivity.
\newblock \emph{Nature}, 483\penalty0 (7391):\penalty0 603--607, 2012.

\bibitem[Beer and Tavazoie(2004)]{GeneExpPredSeq2004}
Michael~A Beer and Saeed Tavazoie.
\newblock Predicting gene expression from sequence.
\newblock \emph{Cell}, 117\penalty0 (2):\penalty0 185--198, 2004.

\bibitem[Benegas et~al.(2022)Benegas, Batra, and Song]{benegas2022dna}
G~Benegas, SS~Batra, and YS~Song.
\newblock Dna language models are powerful zero-shot predictors of non-coding variant effects.
\newblock 2022.

\bibitem[Bengio et~al.(2007)Bengio, Lamblin, Popovici, and Larochelle]{bengio2007sae}
Yoshua Bengio, Pascal Lamblin, Dan Popovici, and Hugo Larochelle.
\newblock Greedy layer-wise training of deep networks.
\newblock In \emph{Advances in neural information processing systems}, pages 153--160, 2007.

\bibitem[Beyreli et~al.(2022)Beyreli, Karakahya, and Cicek]{beyreli2022deepnd}
Ilayda Beyreli, Oguzhan Karakahya, and A~Ercument Cicek.
\newblock Deepnd: Deep multitask learning of gene risk for comorbid neurodevelopmental disorders.
\newblock \emph{Patterns}, 3\penalty0 (7), 2022.

\bibitem[Bohr et~al.(1988)Bohr, Bohr, Brunak, Cotterill, Lautrup, N{\o}rskov, Olsen, and Petersen]{1988earliestProteinSS}
Henrik Bohr, Jakob Bohr, S{\o}ren Brunak, Rodney~MJ Cotterill, Benny Lautrup, Leif N{\o}rskov, Ole~H Olsen, and Steffen~B Petersen.
\newblock Protein secondary structure and homology by neural networks the $\alpha$-helices in rhodopsin.
\newblock \emph{FEBS letters}, 241\penalty0 (1-2):\penalty0 223--228, 1988.

\bibitem[Bonidia et~al.(2021)Bonidia, Domingues, Sanches, and de~Carvalho]{Bonidia2023}
Robson~P Bonidia, Douglas~S Domingues, Danilo~S Sanches, and André C P L~F de~Carvalho.
\newblock {MathFeature: feature extraction package for DNA, RNA and protein sequences based on mathematical descriptors}.
\newblock \emph{Briefings in Bioinformatics}, 23\penalty0 (1):\penalty0 bbab434, 11 2021.
\newblock ISSN 1477-4054.
\newblock \doi{10.1093/bib/bbab434}.
\newblock URL \url{https://doi.org/10.1093/bib/bbab434}.

\bibitem[Bo{\v{z}}a et~al.(2017)Bo{\v{z}}a, Brejov{\'a}, and Vina{\v{r}}]{DeepNano2017}
Vladim{\'\i}r Bo{\v{z}}a, Bro{\v{n}}a Brejov{\'a}, and Tom{\'a}{\v{s}} Vina{\v{r}}.
\newblock Deepnano: Deep recurrent neural networks for base calling in minion nanopore reads.
\newblock \emph{PloS one}, 12\penalty0 (6):\penalty0 e0178751, 2017.

\bibitem[Breda et~al.(2007)Breda, Valadares, de~Souza, and Garratt]{breda2007protein}
Ardala Breda, Napoleao~Fonseca Valadares, Osmar~Norberto de~Souza, and Richard~Charles Garratt.
\newblock Protein structure, modelling and applications.
\newblock 2007.

\bibitem[Brown et~al.(2020)Brown, Mann, Ryder, Subbiah, Kaplan, Dhariwal, Neelakantan, Shyam, Sastry, Askell, et~al.]{gpt-3}
Tom Brown, Benjamin Mann, Nick Ryder, Melanie Subbiah, Jared~D Kaplan, Prafulla Dhariwal, Arvind Neelakantan, Pranav Shyam, Girish Sastry, Amanda Askell, et~al.
\newblock Language models are few-shot learners.
\newblock \emph{Advances in neural information processing systems}, 33:\penalty0 1877--1901, 2020.

\bibitem[Busia et~al.(2016)Busia, Collins, and Jaitly]{busia2016protein}
Akosua Busia, Jasmine Collins, and Navdeep Jaitly.
\newblock Protein secondary structure prediction using deep multi-scale convolutional neural networks and next-step conditioning.
\newblock \emph{arXiv preprint arXiv:1611.01503}, 2016.

\bibitem[Cain et~al.(2011)Cain, Blekhman, Marioni, and Gilad]{2histone2011}
Carolyn~E Cain, Ran Blekhman, John~C Marioni, and Yoav Gilad.
\newblock Gene expression differences among primates are associated with changes in a histone epigenetic modification.
\newblock \emph{Genetics}, 187\penalty0 (4):\penalty0 1225--1234, 2011.

\bibitem[Cang and Wei(2017)]{zixuanTopo2017}
Zixuan Cang and Guo-Wei Wei.
\newblock Topologynet: Topology based deep convolutional and multi-task neural networks for biomolecular property predictions.
\newblock \emph{PLOS Computational Biology}, 13\penalty0 (7):\penalty0 1--27, 07 2017.
\newblock \doi{10.1371/journal.pcbi.1005690}.
\newblock URL \url{https://doi.org/10.1371/journal.pcbi.1005690}.

\bibitem[Cang et~al.(2015)Cang, Mu, Wu, Opron, Xia, and Wei]{cang2015topological}
Zixuan Cang, Lin Mu, Kedi Wu, Kristopher Opron, Kelin Xia, and Guo-Wei Wei.
\newblock A topological approach for protein classification.
\newblock \emph{Molecular Based Mathematical Biology}, 3\penalty0 (1), 2015.

\bibitem[Cao et~al.(2017{\natexlab{a}})Cao, Wu, Ye, and Wang]{cao2017learning}
Jingjun Cao, Zhengli Wu, Wenting Ye, and Haohan Wang.
\newblock Learning functional embedding of genes governed by pair-wised labels.
\newblock In \emph{Computational Intelligence and Applications (ICCIA), 2017 2nd IEEE International Conference on}, pages 397--401. IEEE, 2017{\natexlab{a}}.

\bibitem[Cao et~al.(2015)Cao, Bhattacharya, Adhikari, Li, and Cheng]{cao2015large}
Renzhi Cao, Debswapna Bhattacharya, Badri Adhikari, Jilong Li, and Jianlin Cheng.
\newblock Large-scale model quality assessment for improving protein tertiary structure prediction.
\newblock \emph{Bioinformatics}, 31\penalty0 (12):\penalty0 i116--i123, 2015.

\bibitem[Cao et~al.(2016)Cao, Bhattacharya, Hou, and Cheng]{cao2016deepqa}
Renzhi Cao, Debswapna Bhattacharya, Jie Hou, and Jianlin Cheng.
\newblock Deepqa: improving the estimation of single protein model quality with deep belief networks.
\newblock \emph{BMC bioinformatics}, 17\penalty0 (1):\penalty0 495, 2016.

\bibitem[Cao et~al.(2017{\natexlab{b}})Cao, Freitas, Chan, Sun, Jiang, and Chen]{cao2017prolango}
Renzhi Cao, Colton Freitas, Leong Chan, Miao Sun, Haiqing Jiang, and Zhangxin Chen.
\newblock Prolango: Protein function prediction using neural machine translation based on a recurrent neural network.
\newblock \emph{Molecules}, 22\penalty0 (10):\penalty0 1732, 2017{\natexlab{b}}.

\bibitem[Cao and Zhang(2017)]{cao2017gkmDNN}
Zhen Cao and Shihua Zhang.
\newblock gkm-dnn: efficient prediction using gapped k-mer features and deep neural networks.
\newblock \emph{bioRxiv}, page 170761, 2017.

\bibitem[Castelvecchi(2016)]{openBox}
Davide Castelvecchi.
\newblock Can we open the black box of ai?
\newblock Nature, 2016.
\newblock URL \url{https://www.nature.com/news/can-we-open-the-black-box-of-ai-1.20731?goal=0_997ed6f472-4f78184f7e-154333457&mc_cid=4f78184f7e&mc_eid=74910b9383}.

\bibitem[Chen et~al.(2023)Chen, Cheng, ao~Geng, Li, Zeng, Wang, Gong, Liu, Zeng, Dong, Tang, and Song]{Chen2023xTrimoPGLMU1}
Bo~Chen, Xingyi Cheng, Yangli ao~Geng, Shengyin Li, Xin Zeng, Bo~Wang, Jing Gong, Chiming Liu, Aohan Zeng, Yuxiao Dong, Jie Tang, and Leo~T. Song.
\newblock xtrimopglm: Unified 100b-scale pre-trained transformer for deciphering the language of protein.
\newblock \emph{bioRxiv}, 2023.
\newblock URL \url{https://api.semanticscholar.org/CorpusID:259502990}.

\bibitem[Chen et~al.(2019)Chen, Hou, Shi, Yang, Birchler, and Cheng]{chen2019interpretable}
Chen Chen, Jie Hou, Xiaowen Shi, Hua Yang, James~A Birchler, and Jianlin Cheng.
\newblock Interpretable attention model in transcription factor binding site prediction with deep neural networks.
\newblock \emph{bioRxiv}, page 648691, 2019.

\bibitem[Chen et~al.(2017)Chen, Jacob, and Mairal]{chen2017predicting}
Dexiong Chen, Laurent Jacob, and Julien Mairal.
\newblock Predicting transcription factor binding sites with convolutional kernel networks.
\newblock \emph{bioRxiv}, page 217257, 2017.

\bibitem[Chen et~al.(2016{\natexlab{a}})Chen, Guo, Wang, and Liu]{chen2016comprehensive}
Junjie Chen, Mingyue Guo, Xiaolong Wang, and Bin Liu.
\newblock A comprehensive review and comparison of different computational methods for protein remote homology detection.
\newblock \emph{Briefings in bioinformatics}, page bbw108, 2016{\natexlab{a}}.

\bibitem[Chen et~al.(2016{\natexlab{b}})Chen, Cai, Chen, and Lu]{GeneExpHierachical2016learning}
Lujia Chen, Chunhui Cai, Vicky Chen, and Xinghua Lu.
\newblock Learning a hierarchical representation of the yeast transcriptomic machinery using an autoencoder model.
\newblock \emph{BMC bioinformatics}, 17\penalty0 (1):\penalty0 S9, 2016{\natexlab{b}}.

\bibitem[Chen et~al.(2016{\natexlab{c}})Chen, Li, Narayan, Subramanian, and Xie]{geneExpressionInfer2016}
Yifei Chen, Yi~Li, Rajiv Narayan, Aravind Subramanian, and Xiaohui Xie.
\newblock Gene expression inference with deep learning.
\newblock \emph{Bioinformatics}, 32\penalty0 (12):\penalty0 1832--1839, 2016{\natexlab{c}}.

\bibitem[Cheng et~al.(2011)Cheng, Yan, Yip, Rozowsky, Alexander, Shou, and Gerstein]{SVM2011histone}
Chao Cheng, Koon-Kiu Yan, Kevin~Y Yip, Joel Rozowsky, Roger Alexander, Chong Shou, and Mark Gerstein.
\newblock A statistical framework for modeling gene expression using chromatin features and application to modencode datasets.
\newblock \emph{Genome biology}, 12\penalty0 (2):\penalty0 R15, 2011.

\bibitem[Ching et~al.(2017)Ching, Himmelstein, Beaulieu-Jones, Kalinin, Do, Way, Ferrero, Agapow, Xie, Rosen, et~al.]{opportunitiesObstacles2017}
Travers Ching, Daniel~S Himmelstein, Brett~K Beaulieu-Jones, Alexandr~A Kalinin, Brian~T Do, Gregory~P Way, Enrico Ferrero, Paul-Michael Agapow, Wei Xie, Gail~L Rosen, et~al.
\newblock Opportunities and obstacles for deep learning in biology and medicine.
\newblock \emph{bioRxiv}, page 142760, 2017.

\bibitem[Cho et~al.(2014)Cho, Van~Merri{\"e}nboer, Gulcehre, Bahdanau, Bougares, Schwenk, and Bengio]{gru2014learning}
Kyunghyun Cho, Bart Van~Merri{\"e}nboer, Caglar Gulcehre, Dzmitry Bahdanau, Fethi Bougares, Holger Schwenk, and Yoshua Bengio.
\newblock Learning phrase representations using rnn encoder-decoder for statistical machine translation.
\newblock \emph{arXiv preprint arXiv:1406.1078}, 2014.

\bibitem[Choi and Chae(2020)]{methylome}
Joungmin Choi and Heejoon Chae.
\newblock methcancer-gen: a dna methylome dataset generator for user-specified cancer type based on conditional variational autoencoder.
\newblock \emph{BMC Bioinformatics}, 21, 05 2020.
\newblock \doi{10.1186/s12859-020-3516-8}.

\bibitem[Chu et~al.(2004)Chu, Ghahramani, and Wild]{graphicalSS2004}
Wei Chu, Zoubin Ghahramani, and David~L Wild.
\newblock A graphical model for protein secondary structure prediction.
\newblock In \emph{Proceedings of the twenty-first international conference on Machine learning}, page~21. ACM, 2004.

\bibitem[Cire{\c{s}}an et~al.(2012)Cire{\c{s}}an, Meier, and Schmidhuber]{cirecsan2012transfer}
Dan~C Cire{\c{s}}an, Ueli Meier, and J{\"u}rgen Schmidhuber.
\newblock Transfer learning for latin and chinese characters with deep neural networks.
\newblock In \emph{Neural Networks (IJCNN), The 2012 International Joint Conference on}, pages 1--6. IEEE, 2012.

\bibitem[Cohn et~al.(2018)Cohn, Zuk, and Kaplan]{cohn2018enhancer}
Dikla Cohn, Or~Zuk, and Tommy Kaplan.
\newblock Enhancer identification using transfer and adversarial deep learning of dna sequences.
\newblock \emph{bioRxiv}, page 264200, 2018.

\bibitem[Consortium et~al.(2012)]{encode2012integrated}
ENCODE~Project Consortium et~al.
\newblock An integrated encyclopedia of dna elements in the human genome.
\newblock \emph{Nature}, 489\penalty0 (7414):\penalty0 57--74, 2012.

\bibitem[Cui et~al.(2023)Cui, Wang, Maan, Pang, Luo, and Wang]{cui2023scgpt}
Haotian Cui, Chloe Wang, Hassaan Maan, Kuan Pang, Fengning Luo, and Bo~Wang.
\newblock scgpt: Towards building a foundation model for single-cell multi-omics using generative ai.
\newblock \emph{bioRxiv}, pages 2023--04, 2023.

\bibitem[Dai et~al.(2019)Dai, Yang, Yang, Carbonell, Le, and Salakhutdinov]{Dai2019TransformerXLAL}
Zihang Dai, Zhilin Yang, Yiming Yang, Jaime~G. Carbonell, Quoc~V. Le, and Ruslan Salakhutdinov.
\newblock Transformer-xl: Attentive language models beyond a fixed-length context.
\newblock \emph{ArXiv}, abs/1901.02860, 2019.
\newblock URL \url{https://api.semanticscholar.org/CorpusID:57759363}.

\bibitem[Dalla-Torre et~al.(2023)Dalla-Torre, Gonzalez, Mendoza-Revilla, Carranza, Grzywaczewski, Oteri, Dallago, Trop, Sirelkhatim, Richard, et~al.]{nucleotide-transformer}
Hugo Dalla-Torre, Liam Gonzalez, Javier Mendoza-Revilla, Nicolas~Lopez Carranza, Adam~Henryk Grzywaczewski, Francesco Oteri, Christian Dallago, Evan Trop, Hassan Sirelkhatim, Guillaume Richard, et~al.
\newblock The nucleotide transformer: Building and evaluating robust foundation models for human genomics.
\newblock \emph{bioRxiv}, pages 2023--01, 2023.

\bibitem[Danaee et~al.(2017)Danaee, Ghaeini, and Hendrix]{danaee2017deep}
Padideh Danaee, Reza Ghaeini, and David~A Hendrix.
\newblock A deep learning approach for cancer detection and relevant gene identification.
\newblock In \emph{PACIFIC SYMPOSIUM ON BIOCOMPUTING 2017}, pages 219--229. World Scientific, 2017.

\bibitem[Denas and Taylor(2013)]{denas2013deep}
Olgert Denas and James Taylor.
\newblock Deep modeling of gene expression regulation in an erythropoiesis model.
\newblock In \emph{Representation Learning, ICML Workshop}, 2013.

\bibitem[Devlin et~al.(2018)Devlin, Chang, Lee, and Toutanova]{bert}
Jacob Devlin, Ming-Wei Chang, Kenton Lee, and Kristina Toutanova.
\newblock Bert: Pre-training of deep bidirectional transformers for language understanding.
\newblock \emph{arXiv preprint arXiv:1810.04805}, 2018.

\bibitem[Dincer et~al.(2018)Dincer, Celik, Hiranuma, and Lee]{dincer2018deepprofile}
Ayse~Berceste Dincer, Safiye Celik, Naozumi Hiranuma, and Su-In Lee.
\newblock Deepprofile: Deep learning of patient molecular profiles for precision medicine in acute myeloid leukemia.
\newblock \emph{bioRxiv}, page 278739, 2018.

\bibitem[Dong and Weng(2013)]{3histone2013}
Xianjun Dong and Zhiping Weng.
\newblock The correlation between histone modifications and gene expression.
\newblock 2013.

\bibitem[Dong et~al.(2012)Dong, Greven, Kundaje, Djebali, Brown, Cheng, Gingeras, Gerstein, Guig{\'o}, Birney, et~al.]{RF2012histone}
Xianjun Dong, Melissa~C Greven, Anshul Kundaje, Sarah Djebali, James~B Brown, Chao Cheng, Thomas~R Gingeras, Mark Gerstein, Roderic Guig{\'o}, Ewan Birney, et~al.
\newblock Modeling gene expression using chromatin features in various cellular contexts.
\newblock \emph{Genome biology}, 13\penalty0 (9):\penalty0 R53, 2012.

\bibitem[Du et~al.(2021)Du, Qian, Liu, Ding, Qiu, Yang, and Tang]{Du2021GLMGL}
Zhengxiao Du, Yujie Qian, Xiao Liu, Ming Ding, Jiezhong Qiu, Zhilin Yang, and Jie Tang.
\newblock Glm: General language model pretraining with autoregressive blank infilling.
\newblock In \emph{Annual Meeting of the Association for Computational Linguistics}, 2021.
\newblock URL \url{https://api.semanticscholar.org/CorpusID:247519241}.

\bibitem[Elman(1990)]{elman1990findingRNN}
Jeffrey~L Elman.
\newblock Finding structure in time.
\newblock \emph{Cognitive science}, 14\penalty0 (2):\penalty0 179--211, 1990.

\bibitem[Elnaggar et~al.(2020)Elnaggar, Heinzinger, Dallago, Rehawi, Wang, Jones, Gibbs, Feh{\'e}r, Angerer, Steinegger, Bhowmik, and Rost]{Elnaggar2020ProtTransTC}
Ahmed Elnaggar, Michael Heinzinger, Christian Dallago, Ghalia Rehawi, Yu~Wang, Llion Jones, Tom Gibbs, Tamas~B. Feh{\'e}r, Christoph Angerer, Martin Steinegger, Debsindhu Bhowmik, and Burkhard Rost.
\newblock Prottrans: Towards cracking the language of life’s code through self-supervised deep learning and high performance computing.
\newblock \emph{bioRxiv}, 2020.
\newblock URL \url{https://api.semanticscholar.org/CorpusID:220495861}.

\bibitem[Emanuelsson et~al.(2000)Emanuelsson, Nielsen, Brunak, and Von~Heijne]{emanuelsson2000predicting}
Olof Emanuelsson, Henrik Nielsen, S{\o}ren Brunak, and Gunnar Von~Heijne.
\newblock Predicting subcellular localization of proteins based on their n-terminal amino acid sequence.
\newblock \emph{Journal of molecular biology}, 300\penalty0 (4):\penalty0 1005--1016, 2000.

\bibitem[Faraggi et~al.(2012)Faraggi, Zhang, Yang, Kurgan, and Zhou]{faraggi2012spine}
Eshel Faraggi, Tuo Zhang, Yuedong Yang, Lukasz Kurgan, and Yaoqi Zhou.
\newblock Spine x: improving protein secondary structure prediction by multistep learning coupled with prediction of solvent accessible surface area and backbone torsion angles.
\newblock \emph{Journal of computational chemistry}, 33\penalty0 (3):\penalty0 259--267, 2012.

\bibitem[Fickett and Hatzigeorgiou(1997)]{fickett1997eukaryotic}
James~W Fickett and Artemis~G Hatzigeorgiou.
\newblock Eukaryotic promoter recognition.
\newblock \emph{Genome research}, 7\penalty0 (9):\penalty0 861--878, 1997.

\bibitem[Firpi et~al.(2010)Firpi, Ucar, and Tan]{firpi2010discover}
Hiram~A Firpi, Duygu Ucar, and Kai Tan.
\newblock Discover regulatory dna elements using chromatin signatures and artificial neural network.
\newblock \emph{Bioinformatics}, 26\penalty0 (13):\penalty0 1579--1586, 2010.

\bibitem[Fox et~al.(2013)Fox, Brenner, and Chandonia]{Fox2013SCOPeSC}
Naomi Fox, Steven~E. Brenner, and John-Marc Chandonia.
\newblock Scope: Structural classification of proteins—extended, integrating scop and astral data and classification of new structures.
\newblock \emph{Nucleic Acids Research}, 42:\penalty0 D304 -- D309, 2013.
\newblock URL \url{https://api.semanticscholar.org/CorpusID:14864309}.

\bibitem[Fukushima(1975)]{ae1975cognitron}
Kunihiko Fukushima.
\newblock Cognitron: A self-organizing multilayered neural network.
\newblock \emph{Biological cybernetics}, 20\penalty0 (3-4):\penalty0 121--136, 1975.

\bibitem[Fukushima and Miyake(1982)]{fukushima1982neocognitronCNN}
Kunihiko Fukushima and Sei Miyake.
\newblock Neocognitron: A self-organizing neural network model for a mechanism of visual pattern recognition.
\newblock In \emph{Competition and cooperation in neural nets}, pages 267--285. Springer, 1982.

\bibitem[Ganin et~al.(2016)Ganin, Ustinova, Ajakan, Germain, Larochelle, Laviolette, Marchand, and Lempitsky]{ganin2016domain}
Yaroslav Ganin, Evgeniya Ustinova, Hana Ajakan, Pascal Germain, Hugo Larochelle, Fran{\c{c}}ois Laviolette, Mario Marchand, and Victor Lempitsky.
\newblock Domain-adversarial training of neural networks.
\newblock \emph{The Journal of Machine Learning Research}, 17\penalty0 (1):\penalty0 2096--2030, 2016.

\bibitem[Gao et~al.(2021)Gao, Morini, Salani, Krauson, Chekuri, Sharma, Ragavendran, Erdin, Logan, Li, Dakka, Narasimhan, Zhao, Naryshkin, Trotta, Effenberger, Woll, Gabbeta, Karp, Yu, Johnson, Paquette, Cutting, Talkowski, and Slaugenhaupt]{gao2021}
Dadi Gao, Elisabetta Morini, Monica Salani, Aram~J. Krauson, Anil Chekuri, Neeraj Sharma, Ashok Ragavendran, Serkan Erdin, Emily~M. Logan, Wencheng Li, Amal Dakka, Jana Narasimhan, Xin Zhao, Nikolai Naryshkin, Christopher~R. Trotta, Kerstin~A. Effenberger, Matthew~G. Woll, Vijayalakshmi Gabbeta, Gary Karp, Yong Yu, Graham Johnson, William~D. Paquette, Garry~R. Cutting, Michael~E. Talkowski, and Susan~A. Slaugenhaupt.
\newblock A deep learning approach to identify gene targets of a therapeutic for human splicing disorders.
\newblock \emph{Nature Communications}, 12\penalty0 (1):\penalty0 3332, 2021.
\newblock \doi{10.1038/s41467-021-23663-2}.
\newblock URL \url{https://doi.org/10.1038/s41467-021-23663-2}.

\bibitem[Ghandi et~al.(2014)Ghandi, Lee, Mohammad-Noori, and Beer]{gkmSVM2014}
Mahmoud Ghandi, Dongwon Lee, Morteza Mohammad-Noori, and Michael~A Beer.
\newblock Enhanced regulatory sequence prediction using gapped k-mer features.
\newblock \emph{PLoS computational biology}, 10\penalty0 (7):\penalty0 e1003711, 2014.

\bibitem[Ghotra et~al.(2021)Ghotra, Lee, Tripathy, and Koo]{ghotra2021designing}
Rohan Ghotra, Nicholas~Keone Lee, Rohit Tripathy, and Peter~K Koo.
\newblock Designing interpretable convolution-based hybrid networks for genomics.
\newblock \emph{bioRxiv}, pages 2021--07, 2021.

\bibitem[Gligorijevi{\'c} and Pr{\v{z}}ulj(2015)]{gligorijevic2015methods}
Vladimir Gligorijevi{\'c} and Nata{\v{s}}a Pr{\v{z}}ulj.
\newblock Methods for biological data integration: perspectives and challenges.
\newblock \emph{Journal of the Royal Society Interface}, 12\penalty0 (112):\penalty0 20150571, 2015.

\bibitem[Goodfellow et~al.(2016)Goodfellow, Bengio, and Courville]{goodfellow2016deep}
Ian Goodfellow, Yoshua Bengio, and Aaron Courville.
\newblock \emph{Deep learning}.
\newblock MIT press, 2016.

\bibitem[Gupta et~al.(2015)Gupta, Wang, and Ganapathiraju]{GeneExphhw2015}
Aman Gupta, Haohan Wang, and Madhavi Ganapathiraju.
\newblock Learning structure in gene expression data using deep architectures, with an application to gene clustering.
\newblock In \emph{Bioinformatics and Biomedicine (BIBM), 2015 IEEE International Conference on}, pages 1328--1335. IEEE, 2015.

\bibitem[Hammad et~al.(2023)Hammad, Ghoneim, Mabrouk, and Al-atabany]{hammad_hybrid_2023}
Muhammed~S. Hammad, Vidan~F. Ghoneim, Mai~S. Mabrouk, and Walid~I. Al-atabany.
\newblock A hybrid deep learning approach for {COVID}-19 detection based on genomic image processing techniques.
\newblock \emph{Scientific Reports}, 13\penalty0 (1):\penalty0 4003, March 2023.
\newblock ISSN 2045-2322.
\newblock \doi{10.1038/s41598-023-30941-0}.
\newblock URL \url{https://www.nature.com/articles/s41598-023-30941-0}.

\bibitem[Hao et~al.(2015)Hao, Song, and Storey]{hao2015probabilistic}
Wei Hao, Minsun Song, and John~D Storey.
\newblock Probabilistic models of genetic variation in structured populations applied to global human studies.
\newblock \emph{Bioinformatics}, 32\penalty0 (5):\penalty0 713--721, 2015.

\bibitem[Hao et~al.(2023)Hao, Jing, and Sun]{Hao_Jing_Sun_2023}
Yaru Hao, Xiao-Yuan Jing, and Qixing Sun.
\newblock Cancer survival prediction by learning comprehensive deep feature representation for multiple types of genetic data - bmc bioinformatics, Jun 2023.
\newblock URL \url{https://doi.org/10.1186/s12859-023-05392-z}.

\bibitem[Haque et~al.(2014)Haque, Skinner, and Holder]{haque2014imbalanced}
M~Muksitul Haque, Michael~K Skinner, and Lawrence~B Holder.
\newblock Imbalanced class learning in epigenetics.
\newblock \emph{Journal of Computational Biology}, 21\penalty0 (7):\penalty0 492--507, 2014.

\bibitem[Hawkins and Bod{\'e}n(2006)]{hawkins2006detecting}
John Hawkins and Mikael Bod{\'e}n.
\newblock Detecting and sorting targeting peptides with neural networks and support vector machines.
\newblock \emph{Journal of bioinformatics and computational biology}, 4\penalty0 (01):\penalty0 1--18, 2006.

\bibitem[He and Ma(2013)]{he2013imbalanced}
Haibo He and Yunqian Ma.
\newblock \emph{Imbalanced learning: foundations, algorithms, and applications}.
\newblock John Wiley \& Sons, 2013.

\bibitem[He et~al.(2016)He, Zhang, Ren, and Sun]{ResNet2016deep}
Kaiming He, Xiangyu Zhang, Shaoqing Ren, and Jian Sun.
\newblock Deep residual learning for image recognition.
\newblock In \emph{Proceedings of the IEEE conference on computer vision and pattern recognition}, pages 770--778, 2016.

\bibitem[Hinton and Salakhutdinov(2006)]{dbn2006intro}
Geoffrey~E Hinton and Ruslan~R Salakhutdinov.
\newblock Reducing the dimensionality of data with neural networks.
\newblock \emph{science}, 313\penalty0 (5786):\penalty0 504--507, 2006.

\bibitem[Hinton and Sejnowski(1986)]{rbm1986hinton}
Geoffrey~E Hinton and Terrence~J Sejnowski.
\newblock Learning and releaming in boltzmann machines.
\newblock \emph{Parallel Distrilmted Processing}, 1, 1986.

\bibitem[Ho et~al.(2015)Ho, Hassen, and Le]{rule2015histone}
Bich~Hai Ho, Rania Mohammed~Kotb Hassen, and Ngoc~Tu Le.
\newblock Combinatorial roles of dna methylation and histone modifications on gene expression.
\newblock In \emph{Some Current Advanced Researches on Information and Computer Science in Vietnam}, pages 123--135. Springer, 2015.

\bibitem[Hochreiter and Schmidhuber(1997)]{lstm1997}
Sepp Hochreiter and J{\"u}rgen Schmidhuber.
\newblock Long short-term memory.
\newblock \emph{Neural computation}, 9\penalty0 (8):\penalty0 1735--1780, 1997.

\bibitem[Hochreiter et~al.(2007)Hochreiter, Heusel, and Obermayer]{proFam2007hochreiter}
Sepp Hochreiter, Martin Heusel, and Klaus Obermayer.
\newblock Fast model-based protein homology detection without alignment.
\newblock \emph{Bioinformatics}, 23\penalty0 (14):\penalty0 1728--1736, 2007.

\bibitem[Holley and Karplus(1989)]{basicNN11989holley}
L~Howard Holley and Martin Karplus.
\newblock Protein secondary structure prediction with a neural network.
\newblock \emph{Proceedings of the National Academy of Sciences}, 86\penalty0 (1):\penalty0 152--156, 1989.

\bibitem[Horton and Kanehisa(1992)]{horton1992assessment}
Paul~B Horton and Minoru Kanehisa.
\newblock An assessment of neural network and statistical approaches for prediction of e. coli promoter sites.
\newblock \emph{Nucleic Acids Research}, 20\penalty0 (16):\penalty0 4331--4338, 1992.

\bibitem[Hou et~al.(2017)Hou, Adhikari, and Cheng]{hou2017deepsf}
Jie Hou, Badri Adhikari, and Jianlin Cheng.
\newblock Deepsf: deep convolutional neural network for mapping protein sequences to folds.
\newblock \emph{Bioinformatics}, 2017.

\bibitem[Hou et~al.(2019)Hou, Wu, Cao, and Cheng]{hou2019proteinmulticom}
Jie Hou, Tianqi Wu, Renzhi Cao, and Jianlin Cheng.
\newblock Protein tertiary structure modeling driven by deep learning and contact distance prediction in casp13.
\newblock \emph{Proteins: Structure, Function, and Bioinformatics}, 87\penalty0 (12):\penalty0 1165--1178, 2019.

\bibitem[Hua and Sun(2001)]{SVMss2001novel}
Sujun Hua and Zhirong Sun.
\newblock A novel method of protein secondary structure prediction with high segment overlap measure: support vector machine approach.
\newblock \emph{Journal of molecular biology}, 308\penalty0 (2):\penalty0 397--407, 2001.

\bibitem[Jacobson and Sali(2004)]{jacobson2004comparative}
Matthew Jacobson and Andrej Sali.
\newblock Comparative protein structure modeling and its applications to drug discovery.
\newblock 2004.

\bibitem[JAX(2018)]{jax}
The Jackson~Laboratory JAX.
\newblock Genetics vs. genomics, 2018.
\newblock URL \url{https://www.jax.org/personalized-medicine/precision-medicine-and-you/genetics-vs-genomics}.

\bibitem[Jha et~al.(2017)Jha, Gazzara, and Barash]{jha2017integrative}
Anupama Jha, Matthew~R Gazzara, and Yoseph Barash.
\newblock Integrative deep models for alternative splicing.
\newblock \emph{bioRxiv}, page 104869, 2017.

\bibitem[Ji et~al.(2021)Ji, Zhou, Liu, and Davuluri]{dnabert}
Yanrong Ji, Zhihan Zhou, Han Liu, and Ramana~V Davuluri.
\newblock Dnabert: pre-trained bidirectional encoder representations from transformers model for dna-language in genome.
\newblock \emph{Bioinformatics}, 37\penalty0 (15):\penalty0 2112--2120, 2021.

\bibitem[Jones(1999)]{PSIPRED1999jones}
David~T Jones.
\newblock Protein secondary structure prediction based on position-specific scoring matrices.
\newblock \emph{Journal of molecular biology}, 292\penalty0 (2):\penalty0 195--202, 1999.

\bibitem[Juan-Mateu et~al.(2016)Juan-Mateu, Villate, and Eizirik]{juan2016mechanisms}
Jon{\`a}s Juan-Mateu, Olatz Villate, and D{\'e}cio~L Eizirik.
\newblock Mechanisms in endocrinology: alternative splicing: the new frontier in diabetes research.
\newblock \emph{European journal of endocrinology}, 174\penalty0 (5):\penalty0 R225--R238, 2016.

\bibitem[Jumper et~al.(2021)Jumper, Evans, Pritzel, Green, Figurnov, Ronneberger, Tunyasuvunakool, Bates, {\v{Z}}{\'\i}dek, Potapenko, et~al.]{jumper2021alphafold2}
John Jumper, Richard Evans, Alexander Pritzel, Tim Green, Michael Figurnov, Olaf Ronneberger, Kathryn Tunyasuvunakool, Russ Bates, Augustin {\v{Z}}{\'\i}dek, Anna Potapenko, et~al.
\newblock Highly accurate protein structure prediction with alphafold.
\newblock \emph{Nature}, 596\penalty0 (7873):\penalty0 583--589, 2021.

\bibitem[Kabsch and Sander(1983)]{8state1983kabsch}
Wolfgang Kabsch and Christian Sander.
\newblock Dictionary of protein secondary structure: pattern recognition of hydrogen-bonded and geometrical features.
\newblock \emph{Biopolymers}, 22\penalty0 (12):\penalty0 2577--2637, 1983.

\bibitem[Kang et~al.(2010)Kang, Sul, Zaitlen, Kong, Freimer, Sabatti, Eskin, et~al.]{kang2010variance}
Hyun~Min Kang, Jae~Hoon Sul, Noah~A Zaitlen, Sit-yee Kong, Nelson~B Freimer, Chiara Sabatti, Eleazar Eskin, et~al.
\newblock Variance component model to account for sample structure in genome-wide association studies.
\newblock \emph{Nature genetics}, 42\penalty0 (4):\penalty0 348--354, 2010.

\bibitem[Karli{\'c} et~al.(2010)Karli{\'c}, Chung, Lasserre, Vlahovi{\v{c}}ek, and Vingron]{LR2010histone}
Rosa Karli{\'c}, Ho-Ryun Chung, Julia Lasserre, Kristian Vlahovi{\v{c}}ek, and Martin Vingron.
\newblock Histone modification levels are predictive for gene expression.
\newblock \emph{Proceedings of the National Academy of Sciences}, 107\penalty0 (7):\penalty0 2926--2931, 2010.

\bibitem[Kawai et~al.(2001)Kawai, Shinagawa, Shibata, Yoshino, Itoh, Ishii, Arakawa, Hara, Fukunishi, Konno, et~al.]{FANTOM2001functional}
J~Kawai, A~Shinagawa, K~Shibata, M~Yoshino, M~Itoh, Y~Ishii, T~Arakawa, A~Hara, Y~Fukunishi, H~Konno, et~al.
\newblock Functional annotation of a full-length mouse cdna collection.
\newblock \emph{Nature}, 409\penalty0 (6821):\penalty0 685--690, 2001.

\bibitem[Kelley(2020)]{basenji2}
David~R Kelley.
\newblock Cross-species regulatory sequence activity prediction.
\newblock \emph{PLoS Comput. Biol.}, 16\penalty0 (7):\penalty0 e1008050, July 2020.

\bibitem[Kelley et~al.(2016)Kelley, Snoek, and Rinn]{Basset2016Kelley}
David~R. Kelley, Jasper Snoek, and John~L. Rinn.
\newblock Basset: learning the regulatory code of the accessible genome with deep convolutional neural networks.
\newblock \emph{Genome Res}, 26\penalty0 (7):\penalty0 990--999, Jul 2016.
\newblock ISSN 1088-9051.
\newblock \doi{10.1101/gr.200535.115}.
\newblock URL \url{http://www.ncbi.nlm.nih.gov/pmc/articles/PMC4937568/}.
\newblock 27197224[pmid].

\bibitem[Kelley et~al.(2018)Kelley, Reshef, Bileschi, Belanger, McLean, and Snoek]{basenji1}
David~R Kelley, Yakir~A Reshef, Maxwell Bileschi, David Belanger, Cory~Y McLean, and Jasper Snoek.
\newblock Sequential regulatory activity prediction across chromosomes with convolutional neural networks.
\newblock \emph{Genome Res.}, 28\penalty0 (5):\penalty0 739--750, May 2018.

\bibitem[Kidron et~al.(2005)Kidron, Schechner, and Elad]{sound2005pixels}
Einat Kidron, Yoav~Y Schechner, and Michael Elad.
\newblock Pixels that sound.
\newblock In \emph{Computer Vision and Pattern Recognition, 2005. CVPR 2005. IEEE Computer Society Conference on}, volume~1, pages 88--95. IEEE, 2005.

\bibitem[Kim and Park(2003)]{SVMss20032}
Hyunsoo Kim and Haesun Park.
\newblock Protein secondary structure prediction based on an improved support vector machines approach.
\newblock \emph{Protein Engineering}, 16\penalty0 (8):\penalty0 553--560, 2003.

\bibitem[Kimothi et~al.(2016)Kimothi, Soni, Biyani, and Hogan]{seq2vec2016kimothi}
Dhananjay Kimothi, Akshay Soni, Pravesh Biyani, and James~M Hogan.
\newblock Distributed representations for biological sequence analysis.
\newblock \emph{arXiv preprint arXiv:1608.05949}, 2016.

\bibitem[Kingma and Welling(2013)]{kingma2013auto}
Diederik~P Kingma and Max Welling.
\newblock Auto-encoding variational bayes.
\newblock \emph{arXiv preprint arXiv:1312.6114}, 2013.

\bibitem[Kleftogiannis et~al.(2014)Kleftogiannis, Kalnis, and Bajic]{kleftogiannis2014deep}
Dimitrios Kleftogiannis, Panos Kalnis, and Vladimir~B Bajic.
\newblock Deep: a general computational framework for predicting enhancers.
\newblock \emph{Nucleic acids research}, 43\penalty0 (1):\penalty0 e6--e6, 2014.

\bibitem[Kneller et~al.(1990)Kneller, Cohen, and Langridge]{basicNN31990kneller}
DG~Kneller, FE~Cohen, and R~Langridge.
\newblock Improvements in protein secondary structure prediction by an enhanced neural network.
\newblock \emph{Journal of molecular biology}, 214\penalty0 (1):\penalty0 171--182, 1990.

\bibitem[Kobayashi et~al.(2022)Kobayashi, Cheveralls, Leonetti, and Royer]{cite-key}
Hirofumi Kobayashi, Keith~C. Cheveralls, Manuel~D. Leonetti, and Loic~A. Royer.
\newblock Self-supervised deep learning encodes high-resolution features of protein subcellular localization.
\newblock \emph{Nature Methods}, 19\penalty0 (8):\penalty0 995--1003, 2022.
\newblock \doi{10.1038/s41592-022-01541-z}.
\newblock URL \url{https://doi.org/10.1038/s41592-022-01541-z}.

\bibitem[Krizhevsky et~al.(2012)Krizhevsky, Sutskever, and Hinton]{cnn2012imagenet}
Alex Krizhevsky, Ilya Sutskever, and Geoffrey~E Hinton.
\newblock Imagenet classification with deep convolutional neural networks.
\newblock In \emph{Advances in neural information processing systems}, pages 1097--1105, 2012.

\bibitem[Kryshtafovych and Fidelis(2009)]{kryshtafovych2009protein}
Andriy Kryshtafovych and Krzysztof Fidelis.
\newblock Protein structure prediction and model quality assessment.
\newblock \emph{Drug discovery today}, 14\penalty0 (7-8):\penalty0 386--393, 2009.

\bibitem[Kryshtafovych et~al.(2021)Kryshtafovych, Schwede, Topf, Fidelis, and Moult]{kryshtafovych2021critical}
Andriy Kryshtafovych, Torsten Schwede, Maya Topf, Krzysztof Fidelis, and John Moult.
\newblock Critical assessment of methods of protein structure prediction (casp)—round xiv.
\newblock \emph{Proteins: Structure, Function, and Bioinformatics}, 89\penalty0 (12):\penalty0 1607--1617, 2021.

\bibitem[Kundaje and Zou(2016)]{cs273b}
Anshul Kundaje and James Zou.
\newblock Class lecture, cs 273b: Deep learning in genomics and biomedicine.
\newblock Department of Computer Science, Stanford University, 2016.
\newblock URL \url{https://canvas.stanford.edu/courses/51037}.

\bibitem[Kundaje et~al.(2015)Kundaje, Meuleman, Ernst, Bilenky, Yen, Heravi-Moussavi, Kheradpour, Zhang, Wang, Ziller, et~al.]{Roadmap2015integrative}
Anshul Kundaje, Wouter Meuleman, Jason Ernst, Misha Bilenky, Angela Yen, Alireza Heravi-Moussavi, Pouya Kheradpour, Zhizhuo Zhang, Jianrong Wang, Michael~J Ziller, et~al.
\newblock Integrative analysis of 111 reference human epigenomes.
\newblock \emph{Nature}, 518\penalty0 (7539):\penalty0 317--330, 2015.

\bibitem[Lamb et~al.(2006)Lamb, Crawford, Peck, Modell, Blat, Wrobel, Lerner, Brunet, Subramanian, Ross, et~al.]{CMap2006}
Justin Lamb, Emily~D Crawford, David Peck, Joshua~W Modell, Irene~C Blat, Matthew~J Wrobel, Jim Lerner, Jean-Philippe Brunet, Aravind Subramanian, Kenneth~N Ross, et~al.
\newblock The connectivity map: using gene-expression signatures to connect small molecules, genes, and disease.
\newblock \emph{science}, 313\penalty0 (5795):\penalty0 1929--1935, 2006.

\bibitem[Lanchantin et~al.(2016{\natexlab{a}})Lanchantin, Singh, Lin, and Qi]{deepmotif2016}
Jack Lanchantin, Ritambhara Singh, Zeming Lin, and Yanjun Qi.
\newblock Deep motif: Visualizing genomic sequence classifications.
\newblock \emph{arXiv preprint arXiv:1605.01133}, 2016{\natexlab{a}}.

\bibitem[Lanchantin et~al.(2016{\natexlab{b}})Lanchantin, Singh, Wang, and Qi]{Dashboard}
Jack Lanchantin, Ritambhara Singh, Beilun Wang, and Yanjun Qi.
\newblock Deep gdashboard: Visualizing and understanding genomic sequences using deep neural networks.
\newblock \emph{CoRR}, abs/1608.03644, 2016{\natexlab{b}}.
\newblock URL \url{http://arxiv.org/abs/1608.03644}.

\bibitem[Lander et~al.(2001)Lander, Linton, Birren, Nusbaum, Zody, Baldwin, Devon, Dewar, Doyle, FitzHugh, et~al.]{initialHG2001}
Eric~S Lander, Lauren~M Linton, Bruce Birren, Chad Nusbaum, Michael~C Zody, Jennifer Baldwin, Keri Devon, Ken Dewar, Michael Doyle, William FitzHugh, et~al.
\newblock Initial sequencing and analysis of the human genome.
\newblock \emph{Nature}, 409\penalty0 (6822):\penalty0 860--921, 2001.

\bibitem[Leaver-Fay et~al.(2011)Leaver-Fay, Tyka, Lewis, Lange, Thompson, Jacak, Kaufman, Renfrew, Smith, Sheffler, et~al.]{leaver2011rosetta3}
Andrew Leaver-Fay, Michael Tyka, Steven~M Lewis, Oliver~F Lange, James Thompson, Ron Jacak, Kristian~W Kaufman, P~Douglas Renfrew, Colin~A Smith, Will Sheffler, et~al.
\newblock Rosetta3: an object-oriented software suite for the simulation and design of macromolecules.
\newblock In \emph{Methods in enzymology}, volume 487, pages 545--574. Elsevier, 2011.

\bibitem[LeCun et~al.(1990)LeCun, Boser, Denker, Henderson, Howard, Hubbard, and Jackel]{cnn1990lecun}
Yann LeCun, Bernhard~E Boser, John~S Denker, Donnie Henderson, Richard~E Howard, Wayne~E Hubbard, and Lawrence~D Jackel.
\newblock Handwritten digit recognition with a back-propagation network.
\newblock In \emph{Advances in neural information processing systems}, pages 396--404, 1990.

\bibitem[LeCun et~al.(2015)LeCun, Bengio, and Hinton]{lecun2015deep}
Yann LeCun, Yoshua Bengio, and Geoffrey Hinton.
\newblock Deep learning.
\newblock \emph{Nature}, 521\penalty0 (7553):\penalty0 436--444, 2015.

\bibitem[Lee and Yoon(2015)]{rbm2015boosted}
Taehoon Lee and Sungroh Yoon.
\newblock Boosted categorical restricted boltzmann machine for computational prediction of splice junctions.
\newblock In \emph{International Conference on Machine Learning}, pages 2483--2492, 2015.

\bibitem[Lena et~al.(2012)Lena, Nagata, and Baldi]{DSTnn2012deep}
Pietro~D Lena, Ken Nagata, and Pierre~F Baldi.
\newblock Deep spatio-temporal architectures and learning for protein structure prediction.
\newblock In \emph{Advances in neural information processing systems}, pages 512--520, 2012.

\bibitem[Leung et~al.(2014)Leung, Xiong, Lee, and Frey]{splicingCode2014}
Michael~KK Leung, Hui~Yuan Xiong, Leo~J Lee, and Brendan~J Frey.
\newblock Deep learning of the tissue-regulated splicing code.
\newblock \emph{Bioinformatics}, 30\penalty0 (12):\penalty0 i121--i129, 2014.

\bibitem[Leung et~al.(2016)Leung, Delong, Alipanahi, and Frey]{leung2016machine}
Michael~KK Leung, Andrew Delong, Babak Alipanahi, and Brendan~J Frey.
\newblock Machine learning in genomic medicine: a review of computational problems and data sets.
\newblock \emph{Proceedings of the IEEE}, 104\penalty0 (1):\penalty0 176--197, 2016.

\bibitem[Li et~al.(2015{\natexlab{a}})Li, Chen, Kaye, and Wasserman]{CREs22015identification}
Yifeng Li, Chih-yu Chen, Alice~M Kaye, and Wyeth~W Wasserman.
\newblock The identification of cis-regulatory elements: A review from a machine learning perspective.
\newblock \emph{Biosystems}, 138:\penalty0 6--17, 2015{\natexlab{a}}.

\bibitem[Li et~al.(2015{\natexlab{b}})Li, Chen, and Wasserman]{DFSi2015deep}
Yifeng Li, Chih-Yu Chen, and Wyeth~W Wasserman.
\newblock Deep feature selection: Theory and application to identify enhancers and promoters.
\newblock In \emph{RECOMB}, pages 205--217, 2015{\natexlab{b}}.

\bibitem[Li et~al.(2016{\natexlab{a}})Li, Shi, and Wasserman]{CREs2016genome}
Yifeng Li, Wenqiang Shi, and Wyeth~W Wasserman.
\newblock Genome-wide prediction of cis-regulatory regions using supervised deep learning methods.
\newblock \emph{bioRxiv}, page 041616, 2016{\natexlab{a}}.

\bibitem[Li et~al.(2016{\natexlab{b}})Li, Wu, and Ngom]{li2016review}
Yifeng Li, Fang-Xiang Wu, and Alioune Ngom.
\newblock A review on machine learning principles for multi-view biological data integration.
\newblock \emph{Briefings in bioinformatics}, page bbw113, 2016{\natexlab{b}}.

\bibitem[Li et~al.(2016{\natexlab{c}})Li, Yang, and Zhang]{li2016multiviewL}
Yingming Li, Ming Yang, and Zhongfei Zhang.
\newblock Multi-view representation learning: A survey from shallow methods to deep methods.
\newblock \emph{arXiv preprint arXiv:1610.01206}, 2016{\natexlab{c}}.

\bibitem[Li and Yu(2016)]{cascadSS2016}
Zhen Li and Yizhou Yu.
\newblock Protein secondary structure prediction using cascaded convolutional and recurrent neural networks.
\newblock \emph{arXiv preprint arXiv:1604.07176}, 2016.

\bibitem[Liang et~al.(2015)Liang, Li, Chen, and Zeng]{GeneExpMultimodal2015}
Muxuan Liang, Zhizhong Li, Ting Chen, and Jianyang Zeng.
\newblock Integrative data analysis of multi-platform cancer data with a multimodal deep learning approach.
\newblock \emph{IEEE/ACM Transactions on Computational Biology and Bioinformatics (TCBB)}, 12\penalty0 (4):\penalty0 928--937, 2015.

\bibitem[Liao and Noble(2003)]{liao2003combining}
Li~Liao and William~Stafford Noble.
\newblock Combining pairwise sequence similarity and support vector machines for detecting remote protein evolutionary and structural relationships.
\newblock \emph{Journal of computational biology}, 10\penalty0 (6):\penalty0 857--868, 2003.

\bibitem[Libbrecht(2016)]{libbrecht2016understanding}
Maxwell~Wing Libbrecht.
\newblock \emph{Understanding human genome regulation through entropic graph-based regularization and submodular optimization}.
\newblock PhD thesis, 2016.

\bibitem[Lim et~al.(2009)Lim, Hardy, Bunting, Ma, Peng, Chen, and Shannon]{1histone2009}
Pek~S Lim, Kristine Hardy, Karen~L Bunting, Lina Ma, Kaiman Peng, Xinxin Chen, and Mary~F Shannon.
\newblock Defining the chromatin signature of inducible genes in t cells.
\newblock \emph{Genome biology}, 10\penalty0 (10):\penalty0 R107, 2009.

\bibitem[Lin et~al.(2023)Lin, Akin, Rao, Hie, Zhu, Lu, Smetanin, Verkuil, Kabeli, Shmueli, et~al.]{lin2023evolutionary}
Zeming Lin, Halil Akin, Roshan Rao, Brian Hie, Zhongkai Zhu, Wenting Lu, Nikita Smetanin, Robert Verkuil, Ori Kabeli, Yaniv Shmueli, et~al.
\newblock Evolutionary-scale prediction of atomic-level protein structure with a language model.
\newblock \emph{Science}, 379\penalty0 (6637):\penalty0 1123--1130, 2023.

\bibitem[Lippert et~al.(2011)Lippert, Listgarten, Liu, Kadie, Davidson, and Heckerman]{lippert2011fast}
Christoph Lippert, Jennifer Listgarten, Ying Liu, Carl~M Kadie, Robert~I Davidson, and David Heckerman.
\newblock Fast linear mixed models for genome-wide association studies.
\newblock \emph{Nature methods}, 8\penalty0 (10):\penalty0 833--835, 2011.

\bibitem[Liu et~al.(2017)Liu, Chen, and Li]{liu2017protein}
Bin Liu, Junjie Chen, and Shumin Li.
\newblock Protein remote homology detection based on bidirectional long short-term memory.
\newblock \emph{BMC bioinformatics}, 18\penalty0 (1):\penalty0 443, 2017.

\bibitem[Liu et~al.(2016{\natexlab{a}})Liu, Li, Ren, Bo, and Shu]{liu2016pedla}
Feng Liu, Hao Li, Chao Ren, Xiaochen Bo, and Wenjie Shu.
\newblock Pedla: predicting enhancers with a deep learning-based algorithmic framework.
\newblock \emph{Scientific reports}, 6:\penalty0 28517, 2016{\natexlab{a}}.

\bibitem[Liu et~al.(2022)Liu, Wu, Guo, Hou, and Cheng]{liu2022improvingmulticom2}
Jian Liu, Tianqi Wu, Zhiye Guo, Jie Hou, and Jianlin Cheng.
\newblock Improving protein tertiary structure prediction by deep learning and distance prediction in casp14.
\newblock \emph{Proteins: Structure, Function, and Bioinformatics}, 90\penalty0 (1):\penalty0 58--72, 2022.

\bibitem[Liu et~al.(2016{\natexlab{b}})Liu, Wang, Eickholt, and Wang]{liu2016benchmarking}
Tong Liu, Yiheng Wang, Jesse Eickholt, and Zheng Wang.
\newblock Benchmarking deep networks for predicting residue-specific quality of individual protein models in casp11.
\newblock \emph{Scientific reports}, 6:\penalty0 19301, 2016{\natexlab{b}}.

\bibitem[Lo~Conte et~al.(2000)Lo~Conte, Ailey, Hubbard, Brenner, Murzin, and Chothia]{lo2000scop}
Loredana Lo~Conte, Bart Ailey, Tim~JP Hubbard, Steven~E Brenner, Alexey~G Murzin, and Cyrus Chothia.
\newblock Scop: a structural classification of proteins database.
\newblock \emph{Nucleic acids research}, 28\penalty0 (1):\penalty0 257--259, 2000.

\bibitem[Louizos et~al.(2017)Louizos, Shalit, Mooij, Sontag, Zemel, and Welling]{louizos2017causal}
Christos Louizos, Uri Shalit, Joris~M Mooij, David Sontag, Richard Zemel, and Max Welling.
\newblock Causal effect inference with deep latent-variable models.
\newblock In \emph{Advances in Neural Information Processing Systems}, pages 6449--6459, 2017.

\bibitem[Lu et~al.(2015)Lu, Behbood, Hao, Zuo, Xue, and Zhang]{lu2015transfer}
Jie Lu, Vahid Behbood, Peng Hao, Hua Zuo, Shan Xue, and Guangquan Zhang.
\newblock Transfer learning using computational intelligence: a survey.
\newblock \emph{Knowledge-Based Systems}, 80:\penalty0 14--23, 2015.

\bibitem[Maaten et~al.(2011)Maaten, Welling, and Saul]{hiddenSS2011}
Laurens Maaten, Max Welling, and Lawrence~K Saul.
\newblock Hidden-unit conditional random fields.
\newblock In \emph{International Conference on Artificial Intelligence and Statistics}, pages 479--488, 2011.

\bibitem[Magnan and Baldi(2014)]{rnnSS2014magnan}
Christophe~N Magnan and Pierre Baldi.
\newblock Sspro/accpro 5: almost perfect prediction of protein secondary structure and relative solvent accessibility using profiles, machine learning and structural similarity.
\newblock \emph{Bioinformatics}, 30\penalty0 (18):\penalty0 2592--2597, 2014.

\bibitem[Matis et~al.(1996)Matis, Xu, Shah, Guan, Einstein, Mural, and Uberbacher]{CNNpromoters1996early}
Sherri Matis, Ying Xu, Manesh Shah, Xiaojun Guan, J~Ralph Einstein, Richard Mural, and Edward Uberbacher.
\newblock Detection of rna polymerase ii promoters and polyadenylation sites in human dna sequence.
\newblock \emph{Computers \& chemistry}, 20\penalty0 (1):\penalty0 135--140, 1996.

\bibitem[Mei(2013)]{mei2013probability}
Suyu Mei.
\newblock Probability weighted ensemble transfer learning for predicting interactions between hiv-1 and human proteins.
\newblock \emph{PLoS One}, 8\penalty0 (11):\penalty0 e79606, 2013.

\bibitem[Meinken et~al.(2012)Meinken, Min, et~al.]{meinken2012computational}
John Meinken, Jack Min, et~al.
\newblock Computational prediction of protein subcellular locations in eukaryotes: an experience report.
\newblock \emph{Computational Molecular Biology}, 2\penalty0 (1), 2012.

\bibitem[Mikolov et~al.(2013{\natexlab{a}})Mikolov, Chen, Corrado, and Dean]{mikolov2013efficient}
Tomas Mikolov, Kai Chen, Greg Corrado, and Jeffrey Dean.
\newblock Efficient estimation of word representations in vector space.
\newblock \emph{arXiv preprint arXiv:1301.3781}, 2013{\natexlab{a}}.

\bibitem[Mikolov et~al.(2013{\natexlab{b}})Mikolov, Sutskever, Chen, Corrado, and Dean]{mikolov2013distributed}
Tomas Mikolov, Ilya Sutskever, Kai Chen, Greg~S Corrado, and Jeff Dean.
\newblock Distributed representations of words and phrases and their compositionality.
\newblock In \emph{Advances in neural information processing systems}, pages 3111--3119, 2013{\natexlab{b}}.

\bibitem[Min et~al.(2017)Min, Lee, and Yoon]{min2017deep}
Seonwoo Min, Byunghan Lee, and Sungroh Yoon.
\newblock Deep learning in bioinformatics.
\newblock \emph{Briefings in bioinformatics}, 18\penalty0 (5):\penalty0 851--869, 2017.

\bibitem[Min et~al.(2016)Min, Chen, Chen, and Jiang]{deepenhancer2016}
Xu~Min, Ning Chen, Ting Chen, and Rui Jiang.
\newblock Deepenhancer: Predicting enhancers by convolutional neural networks.
\newblock In \emph{Bioinformatics and Biomedicine (BIBM), 2016 IEEE International Conference on}, pages 637--644. IEEE, 2016.

\bibitem[Mitchell(2017)]{pharmaceutical}
Marit Mitchell.
\newblock Deep genomics applies machine learning to develop new genetic medicines, 2017.
\newblock URL \url{http://news.engineering.utoronto.ca/deep-genomics-applies-machine-learning-develop-new-genetic-medicines/}.

\bibitem[Moon et~al.(2014)Moon, Kim, and Wang]{moon2014multimodal}
Seungwhan Moon, Suyoun Kim, and Haohan Wang.
\newblock Multimodal transfer deep learning with applications in audio-visual recognition.
\newblock \emph{arXiv preprint arXiv:1412.3121}, 2014.

\bibitem[Mooney et~al.(2011)Mooney, Wang, and Pollastri]{mooney2011sclpred}
Catherine Mooney, Yong-Hong Wang, and Gianluca Pollastri.
\newblock Sclpred: protein subcellular localization prediction by n-to-1 neural networks.
\newblock \emph{Bioinformatics}, 27\penalty0 (20):\penalty0 2812--2819, 2011.

\bibitem[Nature(2010)]{GenExp}
Nature.
\newblock Gene expression.
\newblock Nature Education, 2010.
\newblock URL \url{https://www.nature.com/scitable/nated/topicpage/gene-expression-14121669}.

\bibitem[Ng(2017)]{ng2017dna2vec}
Patrick Ng.
\newblock dna2vec: Consistent vector representations of variable-length k-mers.
\newblock \emph{arXiv preprint arXiv:1701.06279}, 2017.

\bibitem[Nguyen et~al.(2023)Nguyen, Poli, Faizi, Thomas, Birch-Sykes, Wornow, Patel, Rabideau, Massaroli, Bengio, Ermon, Baccus, and Ré]{hyenaDNA}
Eric Nguyen, Michael Poli, Marjan Faizi, Armin Thomas, Callum Birch-Sykes, Michael Wornow, Aman Patel, Clayton Rabideau, Stefano Massaroli, Yoshua Bengio, Stefano Ermon, Stephen~A. Baccus, and Chris Ré.
\newblock Hyenadna: Long-range genomic sequence modeling at single nucleotide resolution, 2023.

\bibitem[Nguyen et~al.(2014)Nguyen, Shang, and Xu]{nguyen2014dl}
Son~P Nguyen, Yi~Shang, and Dong Xu.
\newblock Dl-pro: A novel deep learning method for protein model quality assessment.
\newblock In \emph{Neural Networks (IJCNN), 2014 International Joint Conference on}, pages 2071--2078. IEEE, 2014.

\bibitem[Nissen et~al.(2021)Nissen, Johansen, Alles{\o}e, S{\o}nderby, Armenteros, Gr{\o}nbech, Jensen, Nielsen, Petersen, Winther, et~al.]{metagenome}
Jakob~Nybo Nissen, Joachim Johansen, Rosa~Lundbye Alles{\o}e, Casper~Kaae S{\o}nderby, Jose Juan~Almagro Armenteros, Christopher~Heje Gr{\o}nbech, Lars~Juhl Jensen, Henrik~Bj{\o}rn Nielsen, Thomas~Nordahl Petersen, Ole Winther, et~al.
\newblock Improved metagenome binning and assembly using deep variational autoencoders.
\newblock \emph{Nature biotechnology}, 39\penalty0 (5):\penalty0 555--560, 2021.

\bibitem[OpenAI(2023)]{gpt-4}
R~OpenAI.
\newblock Gpt-4 technical report.
\newblock \emph{arXiv}, pages 2303--08774, 2023.

\bibitem[{\"O}ztornaci et~al.(2023){\"O}ztornaci, Syed, Morris, and Ta{\c s}delen]{upsample2023}
R.~Onur {\"O}ztornaci, Hamzah Syed, Andrew~P. Morris, and Bahar Ta{\c s}delen.
\newblock The use of class imbalanced learning methods on ulsam data to predict the case-control status in genome-wide association studies.
\newblock \emph{bioRxiv}, 2023.
\newblock \doi{10.1101/2023.01.05.522884}.
\newblock URL \url{https://www.biorxiv.org/content/early/2023/01/06/2023.01.05.522884}.

\bibitem[Pan and Yang(2010)]{transfer2010survey}
Sinno~Jialin Pan and Qiang Yang.
\newblock A survey on transfer learning.
\newblock \emph{IEEE Transactions on knowledge and data engineering}, 22\penalty0 (10):\penalty0 1345--1359, 2010.

\bibitem[Pan and Shen(2017)]{pan2017rna}
Xiaoyong Pan and Hong-Bin Shen.
\newblock Rna-protein binding motifs mining with a new hybrid deep learning based cross-domain knowledge integration approach.
\newblock \emph{BMC bioinformatics}, 18\penalty0 (1):\penalty0 136, 2017.

\bibitem[Park et~al.(2005)Park, Heo, Kwon, and Chung]{park2005protein}
Dae-Won Park, Hyoung-Sam Heo, Hyuk-Chul Kwon, and Hea-Young Chung.
\newblock Protein function classification based on gene ontology.
\newblock \emph{Information Retrieval Technology}, pages 691--696, 2005.

\bibitem[P{\"a}rnamaa and Parts(2017)]{parnamaa2017Subcellular}
Tanel P{\"a}rnamaa and Leopold Parts.
\newblock Accurate classification of protein subcellular localization from high-throughput microscopy images using deep learning.
\newblock \emph{G3: Genes, Genomes, Genetics}, 7\penalty0 (5):\penalty0 1385--1392, 2017.

\bibitem[Pauling et~al.(1951)Pauling, Corey, and Branson]{3state1951pauling}
Linus Pauling, Robert~B Corey, and Herman~R Branson.
\newblock The structure of proteins: two hydrogen-bonded helical configurations of the polypeptide chain.
\newblock \emph{Proceedings of the National Academy of Sciences}, 37\penalty0 (4):\penalty0 205--211, 1951.

\bibitem[Pearson and Lipman(1988)]{FASTA1988improved}
William~R Pearson and David~J Lipman.
\newblock Improved tools for biological sequence comparison.
\newblock \emph{Proceedings of the National Academy of Sciences}, 85\penalty0 (8):\penalty0 2444--2448, 1988.

\bibitem[Pierleoni et~al.(2006)Pierleoni, Martelli, Fariselli, and Casadio]{pierleoni2006bacello}
Andrea Pierleoni, Pier~Luigi Martelli, Piero Fariselli, and Rita Casadio.
\newblock Bacello: a balanced subcellular localization predictor.
\newblock \emph{Bioinformatics}, 22\penalty0 (14):\penalty0 e408--e416, 2006.

\bibitem[Poli et~al.(2023)Poli, Massaroli, Nguyen, Fu, Dao, Baccus, Bengio, Ermon, and Ré]{hyenaModel}
Michael Poli, Stefano Massaroli, Eric Nguyen, Daniel~Y. Fu, Tri Dao, Stephen Baccus, Yoshua Bengio, Stefano Ermon, and Christopher Ré.
\newblock Hyena hierarchy: Towards larger convolutional language models, 2023.

\bibitem[Pollastri et~al.(2002)Pollastri, Przybylski, Rost, and Baldi]{rnnSS2002pollastri}
Gianluca Pollastri, Darisz Przybylski, Burkhard Rost, and Pierre Baldi.
\newblock Improving the prediction of protein secondary structure in three and eight classes using recurrent neural networks and profiles.
\newblock \emph{Proteins: Structure, Function, and Bioinformatics}, 47\penalty0 (2):\penalty0 228--235, 2002.

\bibitem[Qi et~al.(2010)Qi, Tastan, Carbonell, Klein-Seetharaman, and Weston]{qi2010semi}
Yanjun Qi, Oznur Tastan, Jaime~G Carbonell, Judith Klein-Seetharaman, and Jason Weston.
\newblock Semi-supervised multi-task learning for predicting interactions between hiv-1 and human proteins.
\newblock \emph{Bioinformatics}, 26\penalty0 (18):\penalty0 i645--i652, 2010.

\bibitem[Qian and Sejnowski(1988)]{basicNN21988}
Ning Qian and Terrence~J Sejnowski.
\newblock Predicting the secondary structure of globular proteins using neural network models.
\newblock \emph{Journal of molecular biology}, 202\penalty0 (4):\penalty0 865--884, 1988.

\bibitem[Qin and Feng(2017)]{qin2017imputation}
Qian Qin and Jianxing Feng.
\newblock Imputation for transcription factor binding predictions based on deep learning.
\newblock \emph{PLoS computational biology}, 13\penalty0 (2):\penalty0 e1005403, 2017.

\bibitem[Quang and Xie(2016)]{DanQ2016}
Daniel Quang and Xiaohui Xie.
\newblock Danq: a hybrid convolutional and recurrent deep neural network for quantifying the function of dna sequences.
\newblock \emph{Nucleic Acids Res}, 44\penalty0 (11):\penalty0 e107--e107, Jun 2016.
\newblock ISSN 0305-1048.
\newblock \doi{10.1093/nar/gkw226}.
\newblock URL \url{http://www.ncbi.nlm.nih.gov/pmc/articles/PMC4914104/}.
\newblock 27084946[pmid].

\bibitem[R. et~al.(2021)R., Jain, Kotecha, Pandya, Reddy, E., Varadarajan, Mahanti, and V]{hdnn2021}
Elakkiya R., Deepak~Kumar Jain, Ketan Kotecha, Sharnil Pandya, Sai~Siddhartha Reddy, Rajalakshmi E., Vijayakumar Varadarajan, Aniket Mahanti, and Subramaniyaswamy V.
\newblock Hybrid deep neural network for handling data imbalance in precursor microrna.
\newblock \emph{Frontiers in Public Health}, 9, 2021.
\newblock \doi{10.3389/fpubh.2021.821410}.

\bibitem[Radford et~al.(2018)Radford, Narasimhan, Salimans, Sutskever, et~al.]{gpt-1}
Alec Radford, Karthik Narasimhan, Tim Salimans, Ilya Sutskever, et~al.
\newblock Improving language understanding by generative pre-training.
\newblock 2018.

\bibitem[Radford et~al.(2019)Radford, Wu, Child, Luan, Amodei, Sutskever, et~al.]{gpt-2}
Alec Radford, Jeffrey Wu, Rewon Child, David Luan, Dario Amodei, Ilya Sutskever, et~al.
\newblock Language models are unsupervised multitask learners.
\newblock \emph{OpenAI blog}, 1\penalty0 (8):\penalty0 9, 2019.

\bibitem[Raffel et~al.(2019)Raffel, Shazeer, Roberts, Lee, Narang, Matena, Zhou, Li, and Liu]{Raffel2019ExploringTL}
Colin Raffel, Noam~M. Shazeer, Adam Roberts, Katherine Lee, Sharan Narang, Michael Matena, Yanqi Zhou, Wei Li, and Peter~J. Liu.
\newblock Exploring the limits of transfer learning with a unified text-to-text transformer.
\newblock \emph{ArXiv}, abs/1910.10683, 2019.
\newblock URL \url{https://api.semanticscholar.org/CorpusID:204838007}.

\bibitem[Rampasek and Goldenberg(2017)]{rampasek2017dr}
Ladislav Rampasek and Anna Goldenberg.
\newblock Dr. vae: Drug response variational autoencoder.
\newblock \emph{arXiv preprint arXiv:1706.08203}, 2017.

\bibitem[Rangwala and Karypis(2005)]{rangwala2005profile}
Huzefa Rangwala and George Karypis.
\newblock Profile-based direct kernels for remote homology detection and fold recognition.
\newblock \emph{Bioinformatics}, 21\penalty0 (23):\penalty0 4239--4247, 2005.

\bibitem[Rashid et~al.(2021)Rashid, Shah, Bar-Joseph, and Pandya]{rashid2021dhaka}
Sabrina Rashid, Sohrab Shah, Ziv Bar-Joseph, and Ravi Pandya.
\newblock Dhaka: variational autoencoder for unmasking tumor heterogeneity from single cell genomic data.
\newblock \emph{Bioinformatics}, 37\penalty0 (11):\penalty0 1535--1543, 2021.

\bibitem[Ray et~al.(2012)Ray, Lindahl, and Wallner]{proQ2-2012}
Arjun Ray, Erik Lindahl, and Bj{\"o}rn Wallner.
\newblock Improved model quality assessment using proq2.
\newblock \emph{BMC bioinformatics}, 13\penalty0 (1):\penalty0 224, 2012.

\bibitem[Riesselman et~al.(2017)Riesselman, Ingraham, and Marks]{riesselman2017deepGenerative}
Adam~J Riesselman, John~B Ingraham, and Debora~S Marks.
\newblock Deep generative models of genetic variation capture mutation effects.
\newblock \emph{arXiv preprint arXiv:1712.06527}, 2017.

\bibitem[Rifai et~al.(2011)Rifai, Vincent, Muller, Glorot, and Bengio]{rifai2011cae}
Salah Rifai, Pascal Vincent, Xavier Muller, Xavier Glorot, and Yoshua Bengio.
\newblock Contractive auto-encoders: Explicit invariance during feature extraction.
\newblock In \emph{Proceedings of the 28th International Conference on International Conference on Machine Learning}, pages 833--840. Omnipress, 2011.

\bibitem[Riis and Krogh(1996)]{riis1996structured}
S{\o}ren~Kamaric Riis and Anders Krogh.
\newblock Improving prediction of protein secondary structure using structured neural networks and multiple sequence alignments.
\newblock \emph{Journal of Computational Biology}, 3\penalty0 (1):\penalty0 163--183, 1996.

\bibitem[Rives et~al.(2019)Rives, Goyal, Meier, Guo, Ott, Zitnick, Ma, and Fergus]{Rives2019BiologicalSA}
Alexander Rives, Siddharth Goyal, Joshua Meier, Demi Guo, Myle Ott, C.~Lawrence Zitnick, Jerry Ma, and Rob Fergus.
\newblock Biological structure and function emerge from scaling unsupervised learning to 250 million protein sequences.
\newblock \emph{Proceedings of the National Academy of Sciences of the United States of America}, 118, 2019.
\newblock URL \url{https://api.semanticscholar.org/CorpusID:155162335}.

\bibitem[Rost and Sander(1993{\natexlab{a}})]{70proteinSS1993}
Burkhard Rost and Chris Sander.
\newblock Prediction of protein secondary structure at better than 70\% accuracy.
\newblock \emph{Journal of molecular biology}, 232\penalty0 (2):\penalty0 584--599, 1993{\natexlab{a}}.

\bibitem[Rost and Sander(1993{\natexlab{b}})]{rost1993improved}
Burkhard Rost and Chris Sander.
\newblock Improved prediction of protein secondary structure by use of sequence profiles and neural networks.
\newblock \emph{Proceedings of the National Academy of Sciences}, 90\penalty0 (16):\penalty0 7558--7562, 1993{\natexlab{b}}.

\bibitem[Rost et~al.(1994)Rost, Sander, and Schneider]{rost1994redefining}
Burkhard Rost, Chris Sander, and Reinhard Schneider.
\newblock Redefining the goals of protein secondary structure prediction.
\newblock \emph{Journal of molecular biology}, 235\penalty0 (1):\penalty0 13--26, 1994.

\bibitem[Ruder(2017)]{ruder2017overview}
Sebastian Ruder.
\newblock An overview of multi-task learning in deep neural networks.
\newblock \emph{arXiv preprint arXiv:1706.05098}, 2017.

\bibitem[Ruff and Pappu(2021)]{ruff2021alphafolddisorder}
Kiersten~M Ruff and Rohit~V Pappu.
\newblock Alphafold and implications for intrinsically disordered proteins.
\newblock \emph{Journal of Molecular Biology}, 433\penalty0 (20):\penalty0 167208, 2021.

\bibitem[Rumelhart et~al.(1985)Rumelhart, Hinton, and Williams]{mlp1985bp}
David~E Rumelhart, Geoffrey~E Hinton, and Ronald~J Williams.
\newblock Learning internal representations by error propagation.
\newblock Technical report, California Univ San Diego La Jolla Inst for Cognitive Science, 1985.

\bibitem[Schmidler et~al.(2000)Schmidler, Liu, and Brutlag]{bayesSS2000schmidler}
Scott~C Schmidler, Jun~S Liu, and Douglas~L Brutlag.
\newblock Bayesian segmentation of protein secondary structure.
\newblock \emph{Journal of computational biology}, 7\penalty0 (1-2):\penalty0 233--248, 2000.

\bibitem[Schreiber et~al.(2017)Schreiber, Libbrecht, Bilmes, and Noble]{schreiber2017nucleotide}
Jacob Schreiber, Maxwell Libbrecht, Jeffrey Bilmes, and William Noble.
\newblock Nucleotide sequence and dnasei sensitivity are predictive of 3d chromatin architecture.
\newblock \emph{bioRxiv}, page 103614, 2017.

\bibitem[Schulman et~al.(2022)Schulman, Zoph, Kim, Hilton, Menick, Weng, Uribe, Fedus, Metz, Pokorny, et~al.]{chatgpt}
John Schulman, Barret Zoph, Christina Kim, Jacob Hilton, Jacob Menick, Jiayi Weng, Juan Felipe~Ceron Uribe, Liam Fedus, Luke Metz, Michael Pokorny, et~al.
\newblock Chatgpt: Optimizing language models for dialogue.
\newblock \emph{OpenAI blog}, 2022.

\bibitem[Schuster and Paliwal(1997)]{brnn1997}
Mike Schuster and Kuldip~K Paliwal.
\newblock Bidirectional recurrent neural networks.
\newblock \emph{IEEE Transactions on Signal Processing}, 45\penalty0 (11):\penalty0 2673--2681, 1997.

\bibitem[Schweikert et~al.(2009)Schweikert, R{\"a}tsch, Widmer, and Sch{\"o}lkopf]{schweikert2009empirical}
Gabriele Schweikert, Gunnar R{\"a}tsch, Christian Widmer, and Bernhard Sch{\"o}lkopf.
\newblock An empirical analysis of domain adaptation algorithms for genomic sequence analysis.
\newblock In \emph{Advances in Neural Information Processing Systems}, pages 1433--1440, 2009.

\bibitem[Senior et~al.(2020)Senior, Evans, Jumper, Kirkpatrick, Sifre, Green, Qin, Z{\'i}dek, Nelson, Bridgland, Penedones, Petersen, Simonyan, Crossan, Kohli, Jones, Silver, Kavukcuoglu, and Hassabis]{Senior2020AlphaFold}
Andrew~W. Senior, Richard Evans, John~M. Jumper, James Kirkpatrick, L.~Sifre, Tim Green, Chongli Qin, Augustin Z{\'i}dek, Alexander W.~R. Nelson, Alex Bridgland, Hugo Penedones, Stig Petersen, Karen Simonyan, Steve Crossan, Pushmeet Kohli, David~T. Jones, David Silver, Koray Kavukcuoglu, and Demis Hassabis.
\newblock Improved protein structure prediction using potentials from deep learning.
\newblock \emph{Nature}, 577:\penalty0 706--710, 2020.
\newblock URL \url{https://api.semanticscholar.org/CorpusID:210221987}.

\bibitem[Setty and Leslie(2015)]{setty2015seqgl}
Manu Setty and Christina~S Leslie.
\newblock Seqgl identifies context-dependent binding signals in genome-wide regulatory element maps.
\newblock \emph{PLoS computational biology}, 11\penalty0 (5):\penalty0 e1004271, 2015.

\bibitem[Shao et~al.(2020)Shao, Wang, Sun, Dong, Han, Huang, Zhang, Zhang, and Huang]{multi-task-multi-modal}
Wei Shao, Tongxin Wang, Liang Sun, Tianhan Dong, Zhi Han, Zhi Huang, Jie Zhang, Daoqiang Zhang, and Kun Huang.
\newblock Multi-task multi-modal learning for joint diagnosis and prognosis of human cancers.
\newblock \emph{Medical Image Analysis}, 65:\penalty0 101795, 2020.
\newblock ISSN 1361-8415.
\newblock \doi{https://doi.org/10.1016/j.media.2020.101795}.
\newblock URL \url{https://www.sciencedirect.com/science/article/pii/S1361841520301596}.

\bibitem[Sharifi-Noghabi et~al.(2018)Sharifi-Noghabi, Liu, Erho, Shrestha, Alshalalfa, Davicioni, Collins, and Ester]{sharifi2018deep}
Hossein Sharifi-Noghabi, Yang Liu, Nicholas Erho, Raunak Shrestha, Mohammed Alshalalfa, Elai Davicioni, Colin~C Collins, and Martin Ester.
\newblock Deep genomic signature for early metastasis prediction in prostate cancer.
\newblock \emph{bioRxiv}, page 276055, 2018.

\bibitem[Shatkay et~al.(2007)Shatkay, Höglund, Brady, Blum, Dönnes, and Kohlbacher]{sherLoc2007}
Hagit Shatkay, Annette Höglund, Scott Brady, Torsten Blum, Pierre Dönnes, and Oliver Kohlbacher.
\newblock Sherloc: high-accuracy prediction of protein subcellular localization by integrating text and protein sequence data.
\newblock \emph{Bioinformatics}, 23\penalty0 (11):\penalty0 1410--1417, 2007.
\newblock \doi{10.1093/bioinformatics/btm115}.
\newblock URL \url{+ http://dx.doi.org/10.1093/bioinformatics/btm115}.

\bibitem[Shen et~al.(2020)Shen, Zhang, Han, and Huang]{rnn-attention}
Zhen Shen, Qinhu Zhang, Kyungsook Han, and De-Shuang Huang.
\newblock A deep learning model for rna-protein binding preference prediction based on hierarchical lstm and attention network.
\newblock \emph{IEEE/ACM Trans. Comput. Biol. Bioinformatics}, 19\penalty0 (2):\penalty0 753–762, jul 2020.
\newblock ISSN 1545-5963.
\newblock \doi{10.1109/TCBB.2020.3007544}.
\newblock URL \url{https://doi.org/10.1109/TCBB.2020.3007544}.

\bibitem[Shin et~al.(2017)Shin, Kang, Zhang, and Kihara]{shin2017prediction}
Woong-Hee Shin, Xuejiao Kang, Jian Zhang, and Daisuke Kihara.
\newblock Prediction of local quality of protein structure models considering spatial neighbors in graphical models.
\newblock \emph{Scientific reports}, 7:\penalty0 40629, 2017.

\bibitem[Shrikumar et~al.(2017)Shrikumar, Greenside, and Kundaje]{shrikumar2017reverse}
Avanti Shrikumar, Peyton Greenside, and Anshul Kundaje.
\newblock Reverse-complement parameter sharing improves deep learning models for genomics.
\newblock \emph{bioRxiv}, page 103663, 2017.

\bibitem[Simonyan et~al.(2013)Simonyan, Vedaldi, and Zisserman]{simonyan2013deep}
Karen Simonyan, Andrea Vedaldi, and Andrew Zisserman.
\newblock Deep inside convolutional networks: Visualising image classification models and saliency maps.
\newblock \emph{arXiv preprint arXiv:1312.6034}, 2013.

\bibitem[Singh et~al.(2016{\natexlab{a}})Singh, Lanchantin, Robins, and Qi]{DeepChrome2016}
Ritambhara Singh, Jack Lanchantin, Gabriel Robins, and Yanjun Qi.
\newblock Deepchrome: deep-learning for predicting gene expression from histone modifications.
\newblock \emph{Bioinformatics}, 32\penalty0 (17):\penalty0 i639--i648, 2016{\natexlab{a}}.

\bibitem[Singh et~al.(2017)Singh, Lanchantin, Sekhon, and Qi]{lstm-attention}
Ritambhara Singh, Jack Lanchantin, Arshdeep Sekhon, and Yanjun Qi.
\newblock Attend and predict: Understanding gene regulation by selective attention on chromatin.
\newblock \emph{Advances in neural information processing systems}, 30, 2017.

\bibitem[Singh et~al.(2016{\natexlab{b}})Singh, Yang, Poczos, and Ma]{singh2016EnhancerPromoter}
Shashank Singh, Yang Yang, Barnabas Poczos, and Jian Ma.
\newblock Predicting enhancer-promoter interaction from genomic sequence with deep neural networks.
\newblock \emph{bioRxiv}, page 085241, 2016{\natexlab{b}}.

\bibitem[S{\o}nderby et~al.(2015)S{\o}nderby, S{\o}nderby, Nielsen, and Winther]{sonderby2015convolutionalLSTM}
S{\o}ren~Kaae S{\o}nderby, Casper~Kaae S{\o}nderby, Henrik Nielsen, and Ole Winther.
\newblock Convolutional lstm networks for subcellular localization of proteins.
\newblock In \emph{International Conference on Algorithms for Computational Biology}, pages 68--80. Springer, 2015.

\bibitem[Song et~al.(2015)Song, Hao, and Storey]{song2015testing}
Minsun Song, Wei Hao, and John~D Storey.
\newblock Testing for genetic associations in arbitrarily structured populations.
\newblock \emph{Nature genetics}, 47\penalty0 (5):\penalty0 550, 2015.

\bibitem[Spencer et~al.(2015)Spencer, Eickholt, and Cheng]{2015spencer}
Matt Spencer, Jesse Eickholt, and Jianlin Cheng.
\newblock A deep learning network approach to ab initio protein secondary structure prediction.
\newblock \emph{IEEE/ACM transactions on computational biology and bioinformatics}, 12\penalty0 (1):\penalty0 103--112, 2015.

\bibitem[Steinegger et~al.(2018)Steinegger, Mirdita, and S{\"o}ding]{Steinegger2018ProteinlevelAI_BFD}
Martin Steinegger, Milot Mirdita, and Johannes S{\"o}ding.
\newblock Protein-level assembly increases protein sequence recovery from metagenomic samples manyfold.
\newblock \emph{Nature Methods}, pages 1--4, 2018.
\newblock URL \url{https://api.semanticscholar.org/CorpusID:92596540}.

\bibitem[Stephens et~al.(2015)Stephens, Lee, Faghri, Campbell, Zhai, Efron, Iyer, Schatz, Sinha, and Robinson]{stephens2015big}
Zachary~D Stephens, Skylar~Y Lee, Faraz Faghri, Roy~H Campbell, Chengxiang Zhai, Miles~J Efron, Ravishankar Iyer, Michael~C Schatz, Saurabh Sinha, and Gene~E Robinson.
\newblock Big data: astronomical or genomical?
\newblock \emph{PLoS biology}, 13\penalty0 (7):\penalty0 e1002195, 2015.

\bibitem[Stevens and He(2022)]{stevens2022alphafoldloop}
Amy~O Stevens and Yi~He.
\newblock Benchmarking the accuracy of alphafold 2 in loop structure prediction.
\newblock \emph{Biomolecules}, 12\penalty0 (7):\penalty0 985, 2022.

\bibitem[Stormo(2000)]{stormo2000dna}
Gary~D Stormo.
\newblock Dna binding sites: representation and discovery.
\newblock \emph{Bioinformatics}, 16\penalty0 (1):\penalty0 16--23, 2000.

\bibitem[Sun et~al.(2013)Sun, Muckatira, Yuan, Ji, Newfeld, Kumar, and Ye]{sun2013image}
Qian Sun, Sherin Muckatira, Lei Yuan, Shuiwang Ji, Stuart Newfeld, Sudhir Kumar, and Jieping Ye.
\newblock Image-level and group-level models for drosophila gene expression pattern annotation.
\newblock \emph{BMC bioinformatics}, 14\penalty0 (1):\penalty0 350, 2013.

\bibitem[Suzek et~al.(2014)Suzek, Wang, Huang, McGarvey, and Wu]{Suzek2014UniRefCA}
Baris~E. Suzek, Yuqi Wang, Hongzhan Huang, Peter~B. McGarvey, and Cathy~H. Wu.
\newblock Uniref clusters: a comprehensive and scalable alternative for improving sequence similarity searches.
\newblock \emph{Bioinformatics}, 31:\penalty0 926 -- 932, 2014.
\newblock URL \url{https://api.semanticscholar.org/CorpusID:12423917}.

\bibitem[Svozil et~al.(1997)Svozil, Kvasnicka, and Pospichal]{mlp1997introduction}
Daniel Svozil, Vladimir Kvasnicka, and Jiri Pospichal.
\newblock Introduction to multi-layer feed-forward neural networks.
\newblock \emph{Chemometrics and intelligent laboratory systems}, 39\penalty0 (1):\penalty0 43--62, 1997.

\bibitem[Tan et~al.(2014)Tan, Ung, Cheng, and Greene]{GeneExpADAGE2014}
Jie Tan, Matthew Ung, Chao Cheng, and Casey~S Greene.
\newblock Unsupervised feature construction and knowledge extraction from genome-wide assays of breast cancer with denoising autoencoders.
\newblock In \emph{Pacific Symposium on Biocomputing Co-Chairs}, pages 132--143. World Scientific, 2014.

\bibitem[Tan et~al.(2016)Tan, Hammond, Hogan, and Greene]{ADAGE22016}
Jie Tan, John~H Hammond, Deborah~A Hogan, and Casey~S Greene.
\newblock Adage-based integration of publicly available pseudomonas aeruginosa gene expression data with denoising autoencoders illuminates microbe-host interactions.
\newblock \emph{mSystems}, 1\penalty0 (1):\penalty0 e00025--15, 2016.

\bibitem[Tan et~al.(2017)Tan, Doing, Lewis, Price, Chen, Cady, Perchuk, Laub, Hogan, and Greene]{eADAGE2017}
Jie Tan, Georgia Doing, Kimberley~A Lewis, Courtney~E Price, Kathleen~M Chen, Kyle~C Cady, Barret Perchuk, Michael~T Laub, Deborah~A Hogan, and Casey~S Greene.
\newblock Unsupervised extraction of stable expression signatures from public compendia with eadage.
\newblock \emph{bioRxiv}, page 078659, 2017.

\bibitem[Torng and Altman(2017)]{3dCNN2017Torng}
Wen Torng and Russ~B. Altman.
\newblock 3d deep convolutional neural networks for amino acid environment similarity analysis.
\newblock \emph{BMC Bioinformatics}, 18\penalty0 (1):\penalty0 302, Jun 2017.
\newblock ISSN 1471-2105.
\newblock \doi{10.1186/s12859-017-1702-0}.
\newblock URL \url{https://doi.org/10.1186/s12859-017-1702-0}.

\bibitem[Torracinta and Campagne(2016)]{torracinta2016training}
R{\'e}mi Torracinta and Fabien Campagne.
\newblock Training genotype callers with neural networks.
\newblock \emph{bioRxiv}, page 097469, 2016.

\bibitem[Tran and Blei(2017)]{tran2017implicit}
Dustin Tran and David~M Blei.
\newblock Implicit causal models for genome-wide association studies.
\newblock \emph{arXiv preprint arXiv:1710.10742}, 2017.

\bibitem[Tsimenidis et~al.(2022)Tsimenidis, Vrochidou, and Papakostas]{tsimenidis_omics_2022}
Stefanos Tsimenidis, Eleni Vrochidou, and George~A. Papakostas.
\newblock Omics {Data} and {Data} {Representations} for {Deep} {Learning}-{Based} {Predictive} {Modeling}.
\newblock \emph{International Journal of Molecular Sciences}, 23\penalty0 (20):\penalty0 12272, October 2022.
\newblock ISSN 1422-0067.
\newblock \doi{10.3390/ijms232012272}.
\newblock URL \url{https://www.mdpi.com/1422-0067/23/20/12272}.

\bibitem[Umarov and Solovyev(2017)]{CNNpromoters2017}
Ramzan~Kh Umarov and Victor~V Solovyev.
\newblock Recognition of prokaryotic and eukaryotic promoters using convolutional deep learning neural networks.
\newblock \emph{PloS one}, 12\penalty0 (2):\penalty0 e0171410, 2017.

\bibitem[Urda et~al.(2017)Urda, Montes-Torres, Moreno, Franco, and Jerez]{urda2017deep}
Daniel Urda, Julio Montes-Torres, Fernando Moreno, Leonardo Franco, and Jos{\'e}~M Jerez.
\newblock Deep learning to analyze rna-seq gene expression data.
\newblock In \emph{International Work-Conference on Artificial Neural Networks}, pages 50--59. Springer, 2017.

\bibitem[Uziela et~al.(2016)Uziela, Shu, Wallner, and Elofsson]{ProQ3-2016}
Karolis Uziela, Nanjiang Shu, Bj{\"o}rn Wallner, and Arne Elofsson.
\newblock Proq3: Improved model quality assessments using rosetta energy terms.
\newblock \emph{Scientific reports}, 6:\penalty0 33509, 2016.

\bibitem[Uziela et~al.(2017)Uziela, Men{\'e}ndez~Hurtado, Shu, Wallner, and Elofsson]{uziela2017proq3d}
Karolis Uziela, David Men{\'e}ndez~Hurtado, Nanjiang Shu, Bj{\"o}rn Wallner, and Arne Elofsson.
\newblock Proq3d: improved model quality assessments using deep learning.
\newblock \emph{Bioinformatics}, 33\penalty0 (10):\penalty0 1578--1580, 2017.

\bibitem[Vaswani et~al.(2017)Vaswani, Shazeer, Parmar, Uszkoreit, Jones, Gomez, Kaiser, and Polosukhin]{attention-is-all-you-need}
Ashish Vaswani, Noam Shazeer, Niki Parmar, Jakob Uszkoreit, Llion Jones, Aidan~N Gomez, {\L}ukasz Kaiser, and Illia Polosukhin.
\newblock Attention is all you need.
\newblock \emph{Advances in neural information processing systems}, 30, 2017.

\bibitem[Vincent et~al.(2008)Vincent, Larochelle, Bengio, and Manzagol]{sae2008intro}
Pascal Vincent, Hugo Larochelle, Yoshua Bengio, and Pierre-Antoine Manzagol.
\newblock Extracting and composing robust features with denoising autoencoders.
\newblock In \emph{Proceedings of the 25th international conference on Machine learning}, pages 1096--1103. ACM, 2008.

\bibitem[Wan and Mak(2015)]{wan2015machine}
Shibiao Wan and Man-Wai Mak.
\newblock \emph{Machine learning for protein subcellular localization prediction}.
\newblock Walter de Gruyter GmbH \& Co KG, 2015.

\bibitem[Wang et~al.(2017{\natexlab{a}})Wang, Aragam, and Xing]{wang2017variable}
Haohan Wang, Bryon Aragam, and Eric~P Xing.
\newblock Variable selection in heterogeneous datasets: A truncated-rank sparse linear mixed model with applications to genome-wide association studies.
\newblock \emph{bioRxiv}, page 228106, 2017{\natexlab{a}}.

\bibitem[Wang et~al.(2017{\natexlab{b}})Wang, Gupta, and Xu]{wang2017extracting}
Haohan Wang, Aman Gupta, and Ming Xu.
\newblock Extracting compact representation of knowledge from gene expression data for protein-protein interaction.
\newblock \emph{International Journal of Data Mining and Bioinformatics}, 17\penalty0 (4):\penalty0 279--292, 2017{\natexlab{b}}.

\bibitem[Wang et~al.(2017{\natexlab{c}})Wang, Meghawat, Morency, and Xing]{wang2017select}
Haohan Wang, Aaksha Meghawat, Louis-Philippe Morency, and Eric~P Xing.
\newblock Select-additive learning: Improving generalization in multimodal sentiment analysis.
\newblock In \emph{Multimedia and Expo (ICME), 2017 IEEE International Conference on}, pages 949--954. IEEE, 2017{\natexlab{c}}.

\bibitem[Wang et~al.(2017{\natexlab{d}})Wang, Raj, and Xing]{wang2017origin}
Haohan Wang, Bhiksha Raj, and Eric~P Xing.
\newblock On the origin of deep learning.
\newblock \emph{arXiv preprint arXiv:1702.07800}, 2017{\natexlab{d}}.

\bibitem[Wang et~al.(2022)Wang, Chen, Zhou, Li, Zheng, Wang, Li, and Cui]{wang2022contactlowmsa}
Qin Wang, Jiayang Chen, Yuzhe Zhou, Yu~Li, Liangzhen Zheng, Sheng Wang, Zhen Li, and Shuguang Cui.
\newblock Contact-distil: Boosting low homologous protein contact map prediction by self-supervised distillation.
\newblock In \emph{Proceedings of the AAAI Conference on Artificial Intelligence}, volume~36, pages 4620--4627, 2022.

\bibitem[Wang et~al.(2016{\natexlab{a}})Wang, Peng, Ma, and Xu]{DeepCNF2016Wang}
Sheng Wang, Jian Peng, Jianzhu Ma, and Jinbo Xu.
\newblock Protein secondary structure prediction using deep convolutional neural fields.
\newblock \emph{Sci Rep}, 6:\penalty0 18962, Jan 2016{\natexlab{a}}.
\newblock ISSN 2045-2322.
\newblock \doi{10.1038/srep18962}.
\newblock URL \url{http://www.ncbi.nlm.nih.gov/pmc/articles/PMC4707437/}.
\newblock 26752681[pmid].

\bibitem[Wang et~al.(2017{\natexlab{e}})Wang, Sun, Li, Zhang, and Xu]{ultraDeep2017}
Sheng Wang, Siqi Sun, Zhen Li, Renyu Zhang, and Jinbo Xu.
\newblock Accurate de novo prediction of protein contact map by ultra-deep learning model.
\newblock \emph{PLoS computational biology}, 13\penalty0 (1):\penalty0 e1005324, 2017{\natexlab{e}}.

\bibitem[Wang et~al.(2016{\natexlab{b}})Wang, Liu, Xu, Shi, Zhang, Mo, and Wang]{TopoCpG2016}
Yiheng Wang, Tong Liu, Dong Xu, Huidong Shi, Chaoyang Zhang, Yin-Yuan Mo, and Zheng Wang.
\newblock Predicting dna methylation state of cpg dinucleotide using genome topological features and deep networks.
\newblock 6:\penalty0 19598 EP --, Jan 2016{\natexlab{b}}.
\newblock URL \url{http://dx.doi.org/10.1038/srep19598}.
\newblock Article.

\bibitem[Wang et~al.(2021)Wang, Li, Wang, and Li]{10.1093/bioinformatics/btab185}
Zhiqin Wang, Ruiqing Li, Minghui Wang, and Ao~Li.
\newblock {GPDBN: deep bilinear network integrating both genomic data and pathological images for breast cancer prognosis prediction}.
\newblock \emph{Bioinformatics}, 37\penalty0 (18):\penalty0 2963--2970, 03 2021.
\newblock ISSN 1367-4803.
\newblock \doi{10.1093/bioinformatics/btab185}.
\newblock URL \url{https://doi.org/10.1093/bioinformatics/btab185}.

\bibitem[Ward et~al.(2003)Ward, McGuffin, Buxton, and Jones]{SVMss2003}
Jonathan~J Ward, Liam~J McGuffin, Bernard~F. Buxton, and David~T. Jones.
\newblock Secondary structure prediction with support vector machines.
\newblock \emph{Bioinformatics}, 19\penalty0 (13):\penalty0 1650--1655, 2003.

\bibitem[Wasserman and Sandelin(2004)]{wasserman2004CREs1}
Wyeth~W Wasserman and Albin Sandelin.
\newblock Applied bioinformatics for the identification of regulatory elements.
\newblock \emph{Nature Reviews Genetics}, 5\penalty0 (4):\penalty0 276--287, 2004.

\bibitem[Watson et~al.(1953)Watson, Crick, et~al.]{1stDNAunderstandn1953}
James~D Watson, Francis~HC Crick, et~al.
\newblock Molecular structure of nucleic acids.
\newblock \emph{Nature}, 171\penalty0 (4356):\penalty0 737--738, 1953.

\bibitem[Way and Greene(2017{\natexlab{a}})]{way2017evaluating}
Gregory~P Way and Casey~S Greene.
\newblock Evaluating deep variational autoencoders trained on pan-cancer gene expression.
\newblock \emph{arXiv preprint arXiv:1711.04828}, 2017{\natexlab{a}}.

\bibitem[Way and Greene(2017{\natexlab{b}})]{way2017vaeCancer}
Gregory~P Way and Casey~S Greene.
\newblock Extracting a biologically relevant latent space from cancer transcriptomes with variational autoencoders.
\newblock \emph{bioRxiv}, page 174474, 2017{\natexlab{b}}.

\bibitem[Weinstein et~al.(2013)Weinstein, Collisson, Mills, Shaw, Ozenberger, Ellrott, Shmulevich, Sander, Stuart, Network, et~al.]{weinstein2013cancer}
John~N Weinstein, Eric~A Collisson, Gordon~B Mills, Kenna R~Mills Shaw, Brad~A Ozenberger, Kyle Ellrott, Ilya Shmulevich, Chris Sander, Joshua~M Stuart, Cancer Genome Atlas~Research Network, et~al.
\newblock The cancer genome atlas pan-cancer analysis project.
\newblock \emph{Nature genetics}, 45\penalty0 (10):\penalty0 1113--1120, 2013.

\bibitem[Weiss et~al.(2016)Weiss, Khoshgoftaar, and Wang]{weiss2016survey}
Karl Weiss, Taghi~M Khoshgoftaar, and DingDing Wang.
\newblock A survey of transfer learning.
\newblock \emph{Journal of Big Data}, 3\penalty0 (1):\penalty0 9, 2016.

\bibitem[Weissenow et~al.(2022)Weissenow, Heinzinger, Steinegger, and Rost]{weissenow2022ember3d}
Konstantin Weissenow, Michael Heinzinger, Martin Steinegger, and Burkhard Rost.
\newblock Ultra-fast protein structure prediction to capture effects of sequence variation in mutation movies.
\newblock \emph{bioRxiv}, pages 2022--11, 2022.

\bibitem[Whalen et~al.(2016)Whalen, Truty, and Pollard]{whalen2016enhancer}
Sean Whalen, Rebecca~M Truty, and Katherine~S Pollard.
\newblock Enhancer--promoter interactions are encoded by complex genomic signatures on looping chromatin.
\newblock \emph{Nature genetics}, 48\penalty0 (5):\penalty0 488, 2016.

\bibitem[Widmer and R{\"a}tsch(2012)]{widmer2012multitask}
Christian Widmer and Gunnar R{\"a}tsch.
\newblock Multitask learning in computational biology.
\newblock In \emph{Proceedings of ICML Workshop on Unsupervised and Transfer Learning}, pages 207--216, 2012.

\bibitem[Wu et~al.(2022)Wu, Ding, Wang, Shen, Zhang, Luo, Su, Wu, Xie, Berger, et~al.]{wu2022omegafold}
Ruidong Wu, Fan Ding, Rui Wang, Rui Shen, Xiwen Zhang, Shitong Luo, Chenpeng Su, Zuofan Wu, Qi~Xie, Bonnie Berger, et~al.
\newblock High-resolution de novo structure prediction from primary sequence.
\newblock \emph{BioRxiv}, pages 2022--07, 2022.

\bibitem[Wu et~al.(2021)Wu, Guo, Hou, and Cheng]{wu2021deepdist}
Tianqi Wu, Zhiye Guo, Jie Hou, and Jianlin Cheng.
\newblock Deepdist: real-value inter-residue distance prediction with deep residual convolutional network.
\newblock \emph{BMC bioinformatics}, 22:\penalty0 1--17, 2021.

\bibitem[Wu et~al.(2018)Wu, Wang, Cao, Chen, and Xing]{wu2018fair}
Zhenglin Wu, Haohan Wang, Mingze Cao, Yin Chen, and Eric~P Xing.
\newblock Fair deep learning prediction for healthcare applications with confounder filtering.
\newblock \emph{arXiv preprint arXiv:1803.07276}, 2018.

\bibitem[Xie et~al.(2017)Xie, Wen, Quitadamo, Cheng, and Shi]{xie2017deep}
Rui Xie, Jia Wen, Andrew Quitadamo, Jianlin Cheng, and Xinghua Shi.
\newblock A deep auto-encoder model for gene expression prediction.
\newblock \emph{BMC genomics}, 18\penalty0 (9):\penalty0 845, 2017.

\bibitem[Xiong et~al.(2015)Xiong, Alipanahi, Lee, Bretschneider, Merico, Yuen, Hua, Gueroussov, Najafabadi, Hughes, Morris, Barash, Krainer, Jojic, Scherer, Blencowe, and Frey]{Xiong2015RNAsplicing}
Hui~Y. Xiong, Babak Alipanahi, Leo~J. Lee, Hannes Bretschneider, Daniele Merico, Ryan~KC Yuen, Yimin Hua, Serge Gueroussov, Hamed~S. Najafabadi, Timothy~R. Hughes, Quaid Morris, Yoseph Barash, Adrian~R. Krainer, Nebojsa Jojic, Stephen~W. Scherer, Benjamin~J. Blencowe, and Brendan~J. Frey.
\newblock The human splicing code reveals new insights into the genetic determinants of disease.
\newblock \emph{Science}, 347\penalty0 (6218):\penalty0 1254806--1254806, Jan 2015.
\newblock ISSN 0036-8075.
\newblock \doi{10.1126/science.1254806}.
\newblock URL \url{http://www.ncbi.nlm.nih.gov/pmc/articles/PMC4362528/}.
\newblock 25525159[pmid].

\bibitem[Xiong et~al.(2011)Xiong, Barash, and Frey]{xiong2011bayesian}
Hui~Yuan Xiong, Yoseph Barash, and Brendan~J Frey.
\newblock Bayesian prediction of tissue-regulated splicing using rna sequence and cellular context.
\newblock \emph{Bioinformatics}, 27\penalty0 (18):\penalty0 2554--2562, 2011.

\bibitem[Xu and Yang(2011)]{xu2011survey}
Qian Xu and Qiang Yang.
\newblock A survey of transfer and multitask learning in bioinformatics.
\newblock \emph{Journal of Computing Science and Engineering}, 5\penalty0 (3):\penalty0 257--268, 2011.

\bibitem[Xuan et~al.(2019)Xuan, Cao, Zhang, Kong, and Zhang]{xuan2019dual}
Ping Xuan, Yangkun Cao, Tiangang Zhang, Rui Kong, and Zhaogong Zhang.
\newblock Dual convolutional neural networks with attention mechanisms based method for predicting disease-related lncrna genes.
\newblock \emph{Frontiers in genetics}, 10:\penalty0 416, 2019.

\bibitem[Yang et~al.(2017)Yang, Liu, Ren, Ouyang, Xie, Bo, and Shu]{yang2017biren}
Bite Yang, Feng Liu, Chao Ren, Zhangyi Ouyang, Ziwei Xie, Xiaochen Bo, and Wenjie Shu.
\newblock Biren: predicting enhancers with a deep-learning-based model using the dna sequence alone.
\newblock \emph{Bioinformatics}, 33\penalty0 (13):\penalty0 1930--1936, 2017.

\bibitem[Yang et~al.(2022)Yang, Wang, Wang, Fang, Tang, Huang, Lu, and Yao]{yang2022scbert}
Fan Yang, Wenchuan Wang, Fang Wang, Yuan Fang, Duyu Tang, Junzhou Huang, Hui Lu, and Jianhua Yao.
\newblock scbert as a large-scale pretrained deep language model for cell type annotation of single-cell rna-seq data.
\newblock \emph{Nature Machine Intelligence}, 4\penalty0 (10):\penalty0 852--866, 2022.

\bibitem[Yang et~al.(2014)Yang, Zaitlen, Goddard, Visscher, and Price]{yang2014advantages}
Jian Yang, Noah~A Zaitlen, Michael~E Goddard, Peter~M Visscher, and Alkes~L Price.
\newblock Advantages and pitfalls in the application of mixed-model association methods.
\newblock \emph{Nature genetics}, 46\penalty0 (2):\penalty0 100--106, 2014.

\bibitem[Yang et~al.(2020)Yang, Anishchenko, Park, Peng, Ovchinnikov, and Baker]{yang2020improvedtrrossetta}
Jianyi Yang, Ivan Anishchenko, Hahnbeom Park, Zhenling Peng, Sergey Ovchinnikov, and David Baker.
\newblock Improved protein structure prediction using predicted interresidue orientations.
\newblock \emph{Proceedings of the National Academy of Sciences}, 117\penalty0 (3):\penalty0 1496--1503, 2020.

\bibitem[Yang et~al.(2013)Yang, Soares, Greninger, Edelman, Lightfoot, Forbes, Bindal, Beare, Smith, Thompson, Ramaswamy, Futreal, Haber, Stratton, Benes, McDermott, and Garnett]{GDSCdata}
Wanjuan Yang, Jorge Soares, Patricia Greninger, Elena~J. Edelman, Howard Lightfoot, Simon Forbes, Nidhi Bindal, Dave Beare, James~A. Smith, I.~Richard Thompson, Sridhar Ramaswamy, P.~Andrew Futreal, Daniel~A. Haber, Michael~R. Stratton, Cyril Benes, Ultan McDermott, and Mathew~J. Garnett.
\newblock Genomics of drug sensitivity in cancer (gdsc): a resource for therapeutic biomarker discovery in cancer cells.
\newblock \emph{Nucleic Acids Research}, 41\penalty0 (D1):\penalty0 D955--D961, 2013.
\newblock \doi{10.1093/nar/gks1111}.
\newblock URL \url{+ http://dx.doi.org/10.1093/nar/gks1111}.

\bibitem[Yeung and Ruzzo(2001)]{GeneExpPCA2001}
Ka~Yee Yeung and Walter~L. Ruzzo.
\newblock Principal component analysis for clustering gene expression data.
\newblock \emph{Bioinformatics}, 17\penalty0 (9):\penalty0 763--774, 2001.

\bibitem[Yoon and Kwek(2005)]{yoon2005unsupervised}
Kihoon Yoon and Stephen Kwek.
\newblock An unsupervised learning approach to resolving the data imbalanced issue in supervised learning problems in functional genomics.
\newblock In \emph{Hybrid Intelligent Systems, 2005. HIS'05. Fifth International Conference on}, pages 6--pp. IEEE, 2005.

\bibitem[Yu et~al.(2006)Yu, Pressoir, Briggs, Bi, Yamasaki, Doebley, McMullen, Gaut, Nielsen, Holland, et~al.]{yu2006unified}
Jianming Yu, Gael Pressoir, William~H Briggs, Irie~Vroh Bi, Masanori Yamasaki, John~F Doebley, Michael~D McMullen, Brandon~S Gaut, Dahlia~M Nielsen, James~B Holland, et~al.
\newblock A unified mixed-model method for association mapping that accounts for multiple levels of relatedness.
\newblock \emph{Nature genetics}, 38\penalty0 (2):\penalty0 203--208, 2006.

\bibitem[Yuan et~al.(2007)Yuan, Guo, Shen, and Liu]{GeneExpPredSeq2007re}
Yuan Yuan, Lei Guo, Lei Shen, and Jun~S Liu.
\newblock Predicting gene expression from sequence: a reexamination.
\newblock \emph{PLoS computational biology}, 3\penalty0 (11):\penalty0 e243, 2007.

\bibitem[Zeiler and Fergus(2014)]{zeiler2014visualizing}
Matthew~D Zeiler and Rob Fergus.
\newblock Visualizing and understanding convolutional networks.
\newblock In \emph{European conference on computer vision}, pages 818--833. Springer, 2014.

\bibitem[Zemla et~al.(1999)Zemla, Venclovas, Fidelis, and Rost]{SOV1999zemla}
Adam Zemla, {\v{C}}eslovas Venclovas, Krzysztof Fidelis, and Burkhard Rost.
\newblock A modified definition of sov, a segment-based measure for protein secondary structure prediction assessment.
\newblock \emph{Proteins: Structure, Function, and Bioinformatics}, 34\penalty0 (2):\penalty0 220--223, 1999.

\bibitem[Zeng et~al.(2016)Zeng, Edwards, Liu, and Gifford]{cnnIntro2016}
Haoyang Zeng, Matthew~D Edwards, Ge~Liu, and David~K Gifford.
\newblock Convolutional neural network architectures for predicting dna--protein binding.
\newblock \emph{Bioinformatics}, 32\penalty0 (12):\penalty0 i121--i127, 2016.

\bibitem[Zhang and Shen(2020)]{zhang2020templatethreadai}
Haicang Zhang and Yufeng Shen.
\newblock Template-based prediction of protein structure with deep learning.
\newblock \emph{BMC genomics}, 21\penalty0 (11):\penalty0 1--9, 2020.

\bibitem[Zhang et~al.(2023)Zhang, Chen, Shen, Li, and Sun]{zhang2023enhancingmsaagumenter}
Le~Zhang, Jiayang Chen, Tao Shen, Yu~Li, and Siqi Sun.
\newblock Enhancing the protein tertiary structure prediction by multiple sequence alignment generation.
\newblock \emph{arXiv preprint arXiv:2306.01824}, 2023.

\bibitem[Zhang et~al.(2015)Zhang, Zhou, Hu, Gong, Chen, Cheng, and Zeng]{zhang2015deep}
Sai Zhang, Jingtian Zhou, Hailin Hu, Haipeng Gong, Ligong Chen, Chao Cheng, and Jianyang Zeng.
\newblock A deep learning framework for modeling structural features of rna-binding protein targets.
\newblock \emph{Nucleic acids research}, 44\penalty0 (4):\penalty0 e32--e32, 2015.

\bibitem[Zhang et~al.(2016)Zhang, Li, Zeng, Sun, Kumar, Ye, and Ji]{transferBioIma2016deep}
Wenlu Zhang, Rongjian Li, Tao Zeng, Qian Sun, Sudhir Kumar, Jieping Ye, and Shuiwang Ji.
\newblock Deep model based transfer and multi-task learning for biological image analysis.
\newblock \emph{IEEE Transactions on Big Data}, 2016.

\bibitem[Zhang et~al.(2017)Zhang, An, Hu, Tang, and Yue]{zhang2017hicplus}
Yan Zhang, Lin An, Ming Hu, Jijun Tang, and Feng Yue.
\newblock Hicplus: Resolution enhancement of hi-c interaction heatmap.
\newblock \emph{bioRxiv}, page 112631, 2017.

\bibitem[Zhou and Troyanskaya(2014)]{66Q8ss2014}
Jian Zhou and Olga~G Troyanskaya.
\newblock Deep supervised and convolutional generative stochastic network for protein secondary structure prediction.
\newblock In \emph{International Conference on Machine Learning}, pages 745--753, 2014.

\bibitem[Zhou and Troyanskaya(2015)]{deepsea2015predicting}
Jian Zhou and Olga~G Troyanskaya.
\newblock Predicting effects of noncoding variants with deep learning-based sequence model.
\newblock \emph{Nature methods}, 12\penalty0 (10):\penalty0 931--934, 2015.

\bibitem[Zhou et~al.(2018)Zhou, Theesfeld, Yao, Chen, Wong, and Troyanskaya]{expecto}
Jian Zhou, {Chandra L.} Theesfeld, Kevin Yao, {Kathleen M.} Chen, {Aaron K.} Wong, and {Olga G.} Troyanskaya.
\newblock Deep learning sequence-based ab initio prediction of variant effects on expression and disease risk.
\newblock \emph{Nature Genetics}, 50\penalty0 (8):\penalty0 1171--1179, August 2018.
\newblock ISSN 1061-4036.
\newblock \doi{10.1038/s41588-018-0160-6}.

\bibitem[Zhou et~al.(2023)Zhou, Ji, Li, Dutta, Davuluri, and Liu]{dnabert-2}
Zhihan Zhou, Yanrong Ji, Weijian Li, Pratik Dutta, Ramana Davuluri, and Han Liu.
\newblock Dnabert-2: Efficient foundation model and benchmark for multi-species genome.
\newblock \emph{arXiv preprint arXiv:2306.15006}, 2023.

\bibitem[Zvyagin et~al.(2022)Zvyagin, Brace, Hippe, Deng, Zhang, Bohorquez, Clyde, Kale, Perez-Rivera, Ma, et~al.]{genslm}
Max~T Zvyagin, Alexander Brace, Kyle Hippe, Yuntian Deng, Bin Zhang, Cindy~Orozco Bohorquez, Austin Clyde, Bharat Kale, Danilo Perez-Rivera, Heng Ma, et~al.
\newblock Genslms: Genome-scale language models reveal sars-cov-2 evolutionary dynamics.
\newblock \emph{bioRxiv}, 2022.

\end{thebibliography}

\end{document}